\documentclass[aps,prb,reprint,unsortedaddress, showpacs]{revtex4-1}
\usepackage[utf8]{inputenc}

\usepackage[intlimits]{amsmath}
\usepackage{amsfonts, amssymb}
\usepackage{accents}
\usepackage{graphicx}
\usepackage[novolumeabbr]{unitsdef}
\usepackage{printlen}
\usepackage{subfigure}
\usepackage[version=3]{mhchem}
\usepackage{url}

\DeclareMathOperator{\Tr}{Tr}

\DeclareMathOperator{\diag}{diag}

\renewcommand{\vec}[1]{{\boldsymbol{#1}}}
\newcommand{\mat}[1]{{\boldsymbol{#1}}}
\newcommand{\op}[1]{\hat{\boldsymbol{#1}}}
\newcommand{\mvec}[1]{\tilde{\vec{#1}}}

\newcommand{\mop}[1]{\hat{\tilde{\boldsymbol{#1}}}}

\newcommand{\Nform}{\op{\mathcal{N}}}
\newcommand{\K}{\op{\mathcal{K}}}
\newcommand{\Bext}{\vec{B}^{\mathrm{ext}}}
\newcommand{\Beff}{\vec{B}^{\mathrm{e}}}
\newcommand{\N}{\op{N}}
\newcommand{\cross}{\times}
\newcommand{\J}{\mop{J}}
\newcommand{\m}{\mvec{m}}
\newcommand{\vmu}{\mvec{\mu}}

\newcommand{\cc}{+\text{c.c.}}
\newcommand{\I}{\op{I}}
\newcommand{\rv}{\vec{r}}
\newcommand{\B}{\mop{B}}

\newcommand{\pkappa}{\kappa_\perp}

\newcommand{\muz}{\mu_0}
\newcommand{\Ms}{M_s}
\DeclareMathOperator{\Imag}{Im}
\DeclareMathOperator{\Real}{Re}

\newunit{\dB}{dB}
\newunit{\udot}{dot}
\newunit{\dBperdot}{\dB/\udot}
\begin{document}

\title{Theoretical formalism for collective spin wave edge excitations in arrays
    of dipolarly interacting magnetic nanodots}

\author{Ivan Lisenkov}
\email[]{ivan.lisenkov@phystech.edu}
\affiliation{Department of Physics, Oakland University}
\affiliation{Institute of Radio-engineering and Electronics of RAS, Moscow, Russia}

\author{Vasyl Tyberkevych}
\affiliation{Department of Physics, Oakland University}

\author{Sergey Nikitov}
\affiliation{Institute of Radio-engineering and Electronics of RAS, Moscow, Russia}
\affiliation{Moscow Institute of Physics and Technology (State University), Moscow, Russia}
\affiliation{Saratov State University, Saratov, Russia}

\author{Andrei Slavin}
\affiliation{Department of Physics, Oakland University}

\date{\today}

\pacs{75.30.Ds, 42.70.Qs}

\begin{abstract}
A general theory of spin wave edge modes in semi-infinite and finite periodic arrays of magnetic nanodots existing in a spatially uniform magnetization ground state is developed. The theory is formulated using a formalism of multi-vectors of magnetization dynamics, which allows one to study edge modes in arrays having arbitrarily complex primitive cells and lattice structure. The developed formalism can describe spin wave edge modes localized both at the physical edges of the array and at the internal ``domain walls'' separating the array regions existing in different static magnetization states. Using a perturbation theory, in the framework of the developed formalism it is possible to calculate damping of the edge modes and to describe their excitation by external variable magnetic fields. The theory is illustrated on the following practically important examples: (i) calculation  of  the  FMR absorption in a finite nanodot array having the shape of a right triangle; (ii) calculation of the spectra of nonreciprocal spin wave edge modes, including the modes at the  physical edges of an array and modes at the domain walls inside the array; (iii) study of the influence of the  domain wall modes on the  FMR spectrum of an  array existing in a non-ideal chessboard antiferromagnetic ground state.
\end{abstract}

\maketitle

\section{Introduction}

Novel magnonic~\cite{bib:Demokritov:2012,bib:Krawczyk:2014, bib:Nikitov:2015,bib:Kruglyak:2010a, bib:Serga:2010} and  spintronic~\cite{bib:Sinova:2012, bib:Yongbing:2016} signal processing devices  will require novel dynamically reconfigurable magnetic materials which are able to operate without an external magnetic field. These ``self-biased'' magnetic materials, which can function without heavy and bulky external permanent magnets, will be very attractive for applications in microwave signal processing and magnetic logic.  One of the examples of such self-biased artificial magnetic materials are the arrays of periodically arranged and dipolarly coupled anisotropic magnetic nano-elements, in which the nano-size of the element guarantees its monodomain state, while the shape or/and crystallographic anisotropy determines the definite direction of its static magnetization~\cite{bib:Guslienko:2000, bib:Jorzick:2001,    *bib:Gubbiotti:2004,bib:Galkin:2005,bib:Tacchi:2010, bib:Tacchi:2011, bib:Kruglyak:2010,
    bib:Bondarenko:2010, *bib:Bondarenko:2011, bib:Huber:2011, bib:Keatley:2011,
    bib:Verba:2012,bib:Verba:2013a,bib:Verba:2013, bib:Verba:2013a, bib:Carlotti:2014,
    bib:Khymyn:2014, bib:Kakazei:2015}.
    The static magnetization of each anisotropic nanoelement can have more than one stable direction, which means that an array can exist in several distinct meta-stable static magnetization states\cite{bib:Bondarenko:2011,bib:Khymyn:2014}.  Obviously, the static magnetization state of an array strongly affects the array's dynamic magnetization properties, such as the spectrum of its spin wave excitations and characteristics of the array's interaction with external electromagnetic waves\cite{bib:Verba:2012,bib:Verba:2013,bib:Lisenkov:2015}. Moreover, the static magnetization state of an array can be dynamically switched between several metastable configurations by application of a short pulse of an external bias magnetic field of a particular direction and duration~\cite{bib:Verba:2012a, bib:Verba:2013a}. Thus, a magnetic metamaterial based on an array of dipolarly coupled magnetic nanoelements is, indeed, \emph{dynamically reconfigurable}, as its static and dynamic magnetic and electromagnetic properties can be  substantially altered without any changes to its physical structure or composition.

Although the recent progress in the electron and ion beam lithography has made possible the fabrication of arrays of magnetic nanoelements (nanodots) on a large scale~\cite{bib:Gubbiotti:2004, bib:Shaw:2007, bib:Tacchi:2010,bib:Kruglyak:2010, bib:Tacchi:2011,bib:Wang:2011, bib:Semenova:2013, bib:Carlotti:2014, bib:Kakazei:2015}, the fabrication of arrays of elements with high aspect ratios (height substantially larger than the radius), which is necessary for the effective dipolar interaction between the elements~\cite{bib:Kakazei:2015}, still remains a challenging technological problem.   Nonetheless, we believe that materials with uniaxial crystallographic anisotropy\cite{bib:Weller:2000}, like multi-layered composites of \ce{CoPt}, \ce{Co3Pt}, \ce{FePt}, \ce{CoPd}~\cite{bib:Schmool:2008, *bib:Bandiera:2012, *bib:Pal:2014}, which exhibit strongly perpendicular static magnetization in the absence of a bias magnetic field, will make possible the physical realization of the magnetic nanodot arrays for signal processing applications in the near future.

Integration of nano-structured magnetic metamaterials with modern CMOS electronics will require miniaturization, and, therefore, the presence of edge effects in relatively small pieces if magnetic metamaterials will play an increasingly important role with the progress in the systems miniaturization~\cite{bib:Puszkarski:1973, bib:Jorzick:2002, bib:Lisenkov:2013, bib:Lisenkov:2014}. Thus, the properties of the edge excitations in such systems should be well-understood.
In one of our previous works~\cite{bib:Lisenkov:2014} we calculated the spectrum  of collective spin wave edge modes in a periodic  dipolarly coupled nanodot array having one dot per a primitive cell.  However, for an array of magnetic nanodots to have such interesting and unusual properties as non-reciprocity of a spin wave spectra\cite{bib:Verba:2013} or non-trivial topological properties of a spin wave pass-bands~\cite{bib:Shindou:2013, *bib:Shindou:2013a, bib:Chisnell:2015}, it is necessary to have a \emph{complex }primitive cell, e.g.  a primitive cell containing several elements of a different kind or several similar elements having different orientations of their static magnetization.

The idea to use analytical methods capable of describing magnetization dynamics in arrays of magnetic elements goes back to the time when the first ferrite computer memory arrays were developed. One of the first attempts to calculate the distribution of the static demagnetization magnetic fields in an infinite 3D array of magnetic spheres arranged in a periodic lattice was undertaken by Kacz\'{e}r and Murtinov\'{a}~\cite{bib:Kaczer:1974}, where they used a Fourier expansion of a magnetization distribution across the 3D lattice,  and solved the Poisson equation in the reciprocal space. Later, several theoretical approaches were developed to describe spin-wave excitations in systems where magnetic properties are spatially periodic. The developed approaches include the method of plane wave expansion (PWE)~\cite{bib:Vasseur:1996, bib:Nikitov:2001, bib:Krawczyk:2008, bib:Kostylev:2008}, which was adopted from the theory of periodic dielectric~\cite{bib:Ho:1990} and acoustic ~\cite{bib:Kushwaha:1993} structures, the dynamic matrix method~\cite{bib:Grimsditch:2004, bib:Tacchi:2011, bib:Malago:2015, bib:Buijnsters:2014}, the transfer matrix method~\cite{bib:Kruglyak:2006, bib:Urmancheev:2016}, the multiple scattering theory~\cite{bib:Lisenkov:2013}, and  several other.  The above mentioned methods have proven their applicability and convenience for the analysis of \emph{infinite} periodic magnetic systems.

The attempts to handle problems of spin wave excitations in \emph{finite periodic structures with boundaries}, however, faced significant difficulties. An accurate treatment of the finite magnetic systems require: (i) taking into account the boundary conditions at the system's edges, and (ii)taking into account the demagnetization field for all the periodic structure.  This demagnetization field is shape-dependent (and not lattice-dependent), and, also, is non-uniform across the structure~\cite{bib:Lisenkov:2014, bib:Jorzick:2002}. Several attempts were undertaken to treat finite magnetic periodic structures, either by using the PWE methods~\cite{bib:Klos:2011, bib:Lisenkov:2015a} or by developing dedicated methods for special cases ~\cite{bib:Shindou:2013a}.   However,  since the spatial Fourier harmonics of the magnetization and magnetic field do not satisfy the boundary conditions  at the edges of the finite magnetic structure automatically, in finite systems with boundaries the PWE-based methods lose their simplicity and elegance.

Dynamics of spin wave excitations in a finite array of dipolarly coupled magnetic dots, in principle, can be simulated using one of the available micro-magnetic numerical techniques~\cite{bib:Donahue:1999, *bib:Fischbacher:2007, *bib:Vansteenkiste:2014, bib:Buijnsters:2014, bib:Grimsditch:2004}.  However, using only the results of micromagnetic simulations, that provide the frequencies and profiles of the system's eigenmodes, it is, sometimes, difficult to extract the symmetry properties of the system ~\cite{bib:Verba:2013, bib:Guslienko:2008} or to describe the system's behavior in a critical regime, when the stable static state of the system is changed  ~\cite{bib:Verba:2012a, bib:Verba:2013a}. Also,  direct micromagnetic simulations could be computationally intensive, while their results  would be neither scalable nor reusable, since the change of any of the system parameters, such as the quasi-stable state of the array's static magnetization or/and the material properties of the array's elements, would require a complete rerun of the entire simulation.

Thus, below, in our attempt to describe the general properties of collective edge modes in magnetic dot arrays, we decided to use approximate analytical methods based on the Fourier transform of the mutual demagnetization tensor of individual array's elements\cite{bib:Kaczer:1974,bib:Beleggia:2003, bib:Beleggia:2004, bib:Verba:2012, bib:Lisenkov:2014} and an operator form of the linearized Landau-Lifshitz equation~\cite{bib:Buijnsters:2014}. The analytical approach developed by Verba {\em et al.}~\cite{bib:Verba:2012} calculates the spin wave spectra in spatially infinite periodic arrays of magnetic nanodots using the fundamental tensor $\op{F}_\vec{k}$ of the array. This tensor contains all the information about the array (including the lattice symmetry and other geometrical properties of the array's element) that is necessary for the calculation of the spin wave spectrum of the array. The symmetry properties of the fundamental tensor $\op{F}_\vec{k}$ can be evaluated analytically, providing an opportunity to check the symmetric features of the  spin wave spectrum, such as non-reciprocity, analytically without preforming time-consuming numerical calculations~\cite{bib:Verba:2013}. The operator from of the linearized Landau-Lifshitz equation developed by Buijnsters {\em et al.}~\cite{bib:Buijnsters:2014} allows one to reduce the calculation the spectrum of collective spin wave excitations of the array to solving a generalized Hermitian eigenvalue problem.

In this paper, we will generalize the analytical method developed in~[\onlinecite{bib:Verba:2012} and~\onlinecite{bib:Lisenkov:2014}] to include the possibility of theoretical analysis of the spectrum of collective edge spin wave modes in a \emph{semi-infinite} array of magnetic nanodots with a \emph{complex} primitive cell. Our method uses the ``macrospin'' approximation, which assumes that each individual magnetic dot has a spatially uniform distribution of magnetization. In general, the modes having a uniform spatial distribution are expected to be dominant in magnetic nanodot arrays\cite{bib:Kruglyak:2010, bib:Tacchi:2010},however, one can easily extend the theoretical formalism developed below to handle the cases with non-uniform mode profiles simply by adjusting the form of the mutual demagnetization tensor.  A brief description of the possible extension of our formalism  beyond the macrospin approximation is presented in Appendix ~\ref{app:ext}. The formalism presented below can describe both the spin wave modes localized at the physical edge of a semi-infinite array of magnetic dots and the spin wave excitations localized at the array's internal boundaries formed e.g by the domain walls separating the regions existing in different meta-stable states of static magnetization. The latter type of the localized spin wave excitations is analogous to the  Winter magnons~\cite{bib:Winter:1961,*bib:Borovik:1977} existing near the domain walls in a continuous ferromagnetic medium.

Another feature of the analytical formalism developed below is that it makes possible to use a conventional a perturbation theory to find the damping rates of the edge spin wave modes and their coupling to spatially uniform external magnetic fields. In particular, this technique allows one to calculate the ferromagnetic resonance (FMR) absorption spectra of \emph{finite} magnetic nanodot arrays, in which the influence of the edge spin wave modes could be quite significant.

The paper is organized as follows: Sec.~\ref{sec:form} gives a basic description of the dipolar interaction between the nanodots in an array and formulates the general equations that are used in the further spectral calculations; Sec.~\ref{sec:mult} introduces a \emph{multi-vector} formalism for magnetic dot arrays having a complex primitive cell and presents equations necessary for the calculations of the spectra of  bulk and edge spin wave excitations in such arrays; Sec.~\ref{sec:fmr} formulates a perturbative technique used to calculate a response of a finite magnetic dot array on an externally applied microwave magnetic field; Sec.~\ref{sec:examples:pecu} is devoted to the discussion of the specific features of the numerical solutions for the equations derived in~Sec.~\ref{sec:mult}, while Sec.~\ref{sec:examples} deals with the several examples showing the applications of the developed analytical theory to real physical systems:(i) calculation  of  the  FMR absorption spectrum in a finite magnetic nanodot array having the shape of a right triangle; (ii) calculation of nonreciprocal spin wave spectra of edge modes, including the modes at the  physical edges of an array and the modes localized at the domain walls inside an array;(iii) study of the influence of the  domain wall modes on the  FMR spectrum of an  array existing in a non-ideal chessboard antiferromagnetic ground state; Sec.~\ref{sec:conclusions} formulates the conclusions of our work. The developed theoretical  formalism is implemented in a computer program which is available to general public at~[\onlinecite{bib:Lisenkov:program}].

\section{Basic formulation}
\label{sec:form}
\subsection{Mutual demagnetizing tensor with uniaxial anisotropy}
Let us consider an array of dipolarly coupled magnetic nanodots. For the sake of mathematical simplicity we consider dots having the identical shape and saturation magnetization $M_s$. The dots may, however, have different values of the uniaxial crystalline anisotropy. The spatial position of a dot denoted by the index $i$ and having the static magnetization vector $\vec{M}_i$ is determined by the position vector $\rv_i$. For a dot having the index $i$ the effective magnetic field acting on the dot can be written as:
\begin{equation}
    \Beff_i = \Bext_i  + \dfrac{2K^a_i}{\mu_0 M_s^2}\vec{n}_i(\vec{n}_i\cdot\vec{M}_i)
    - \mu_0\sum_j \Nform_{ij} \cdot\vec{M}_j,
    \label{eq:form:magn_field}
\end{equation}
where $\Bext_i$ is the external magnetic field, $K^a_i$ is the energy of the first order uniaxial anisotropy~\cite{bib:Stancil:2009, bib:Gurevich:1996} of the $i$-th dot, $\vec{n}_i$ is the unit vector directed along the anisotropy axis, and $\Nform_{ij} = \Nform(\rv_i-\rv_j)$ is the mutual demagnetization tensor between the dots $i$ and $j$. This tensor is defined by the dot shape and the interdot
distance~\cite{bib:Beleggia:2004,bib:Verba:2012}. The second term in the right-hand-side part of
\eqref{eq:form:magn_field} is the anisotropy of the $i$-th dot. This field can be represented in terms of a tensor $\K_i$ having the following form~\cite{bib:Stancil:2009}:
\begin{equation}
    \K_i = -\dfrac{2K^a_i}{\mu_0 M_s^2}\, \vec{n}_i\otimes\vec{n}_i,
\end{equation}
where $\otimes$ is the direct vector product.

The tensors $\Nform_{ij}$ and $\K_i$ enter~\eqref{eq:form:magn_field} in a similar way, so that it is possible to introduce an effective demagnetization tensor:
\begin{equation}
    \N_{ij} = \Nform_{ij} + \delta_{ij}\K_j,
\end{equation}
and to rewrite~\eqref{eq:form:magn_field} in the form:
\begin{equation}
    \Beff_i = \Bext_i - \mu_0\sum_j \N_{ij}\cdot\vec{M}_j,
    \label{eq:form:magn_field_Neff}
\end{equation}
which formally coincides with equation~(3.4) in Ref.~[\onlinecite{bib:Verba:2012}] for dipolarly coupled \emph{isotropic} magnetic nanodots.  The introduction of the effective demagnetization tensor $\op{N}_{ij}$ in the form (3) allows us to use some results from the previous works~\cite{bib:Verba:2012,bib:Verba:2013} in the case of anisotropic magnetic dots, since the symmetric properties of the tensor $\N_{ij}$ remain the same, with the exception that $\Tr(\N_{ii})= 1 - 2K^a_i/(\muz\Ms^2)$.

Although the above described procedure cannot be applied to some types of the crystalline anisotropy (e.g.~cubic), in most practical cases the uniaxial crystalline and shape (defined by the tensor $\Nform_{ij}$) anisotropies of a magnetic nanoelement are dominant. Then, the effective demagnetization tensor $\op{N}_{ij}$ gives an adequate
description of the magnetic properties of a magnetic dot array.

We would also like to note, that, although the above presented approach cannot be directly used  for the arrays of dots having different shapes and volumes, one can approximate such arrays by arrays of dots having the same shape and volume, but different uniaxial anisotropies, and the above presented approach will allow one to obtain results that are qualitatively correct\cite{bib:Verba:2013}.

\subsection{Equation of motion in the operator form}
Spin wave dynamics for an array of magnetic nanodots is described by the Landau-Lifshitz equation for each dot:
\begin{equation}
    \dfrac{d\vec{M}_i}{dt} = \gamma(\Beff_i\cross\vec{M}_i),
    \label{eq:form:LL}
\end{equation}
where $\gamma/2\pi \approx  28\ilu{\GHz/\tesla}$ is the modulus of the gyromagnetic ratio. In this work we are interested in spin waves with small precession angles. To
linearize \eqref{eq:form:LL} we decompose the magnetization vector into the static and dynamic parts:
\begin{equation}
    \vec{M}_i = M_s(\vec{\mu}_i + \vec{m}_i) + O(|\vec{m}_i|^2),
    \label{eq:form:M_expansion}
\end{equation}
where $\vec{m}_i$ is a small dimensionless deviation of the magnetization of $i$-th dot from the equilibrium. The direction of the static magnetization of a dot is defined by the unit vector
$\vec{\mu}_i$. Since the length of the magnetization vector $\vec{M}_i$ is conserved, the vectors $\vec{\mu}_i$ and $\vec{m}_i$ are orthogonal to each other:
\begin{equation}
    \vec{\mu}_i\cdot\vec{m}_i = 0.
    \label{eq:form:m_mu_otrh}
\end{equation}
In equilibrium, the effective magnetic field acting on each dot is parallel to the static magnetization of the dot:
\begin{equation}
    \Beff_i = B_i\vec{\mu}_i,
\end{equation}
where $B_i$ is the scalar internal field in the $i$-th dot.
To solve the equation of motion one should plug \eqref{eq:form:magn_field_Neff}
and~\eqref{eq:form:M_expansion} into \eqref{eq:form:LL} using condition \eqref{eq:form:m_mu_otrh},
and retain only the terms linear in $\vec{m}_i$. Spin-wave eigenmodes are the harmonic solutions of the linearized equations, $\vec{m}_i(t) = \vec{m}_{i}e^{-i\omega t}\cc$, where $\omega$ is the frequency of the mode and $\vec{m}_{i}$ is the complex mode profile.  Using this decomposition one can linearize \eqref{eq:form:LL} and split it into two equations, for static and dynamic parts of the magnetization, respectively:
\begin{align}
    B_i\vec{\mu}_i &= \Bext_i - \mu_0 M_s\sum_j\op{N}_{ij}\cdot\vec{\mu}_j,\label{eq:form:sum_static}\\
    -i\omega\vec{m}_{i} &=
    \vec{\mu}_i\times\sum_j\op{\Omega}_{ij}\cdot\vec{m}_{j},\label{eq:form:sum_dynamic}
\end{align}
where
\begin{equation}
\op\Omega_{ij} = \gamma B_i\delta_{ij}\op I + \gamma \mu_0 M_s \op N_{ij},
\label{eq:form:omega_general}
\end{equation}
 and $\op I$ is the identity matrix. These equations completely define the behavior of the nanodot array. Solutions of these equations for a finite aperiodic array and an infinite periodic array have been discussed previously in~[\onlinecite{bib:Verba:2012}].

Equation~\eqref{eq:form:sum_dynamic} contains a cross product operation, which is not convenient for the further analysis. One can eliminate the cross product by formally replacing it by the operator $\op{J}_i=\mat{e}\cdot\vec{\mu}_i$ where $\mat{e}$ is the Levi-Civita symbol~\cite{bib:Riley:2006}.
It can be shown by direct substitution that the tensor $\op{J}_i$ has the following properties:
\begin{equation}
    \begin{aligned}
        \op{J}_i^T &= - \op{J}_i\\
        -\op{J}_i^2 =\op{P}_i&= \I -\vec{\mu}_i\otimes\vec{\mu}_i,\\
        \op{P}_i\cdot \vec{m}_i &= \vec{m}_i.
    \end{aligned}
    \label{eq:form:J_prop}
\end{equation}
Here, $\op{P}_i$ is the projection operator to the plane that is
perpendicular to the vector $\vec{\mu}_i$.

Substituting the operator $\op{J}_i$ for the cross product in \eqref{eq:form:sum_dynamic}, multiplying the resulting expression by $\op{J}_i$, and using the properties in \eqref{eq:form:J_prop} one
obtains the following equation:
\begin{equation}
    -i\omega\op{J}_i\cdot\vec{m}_{i} = \sum_{j}\op{\Omega}'_{ij}\cdot\vec{m}_{j},
    \label{eq:form:main_dynamic}
\end{equation}
where
\begin{equation}
    \op\Omega'_{ij} = \op{P}_i\cdot\op\Omega_{ij}\cdot\op{P}_j.
    \label{eq:form:projection}
\end{equation}
Equation~\eqref{eq:form:main_dynamic} is a generalized eigenvalue problem, where $\op{J}_i$ is
an antisymmetric matrix and $\op\Omega'_{ij}$ is a real symmetric matrix (in the sense $\op{\Omega}'_{ij}
= {\op{\Omega}'}^T_{ij} $).  Using
the properties of the tensors $\op{J}_i$ and $\op{\Omega}'_{ij}$ one can immediately show that the eigenfrequencies $\omega$ are real for all stable static magnetic configurations of the
array~\cite{bib:Verba:2012}. We note, that the similar operator approach was successfully used to study Goldstone's modes in spin exchange-coupled systems~\cite{bib:Buijnsters:2014}.

The vector $\vec{m}_i$ is a three-dimensional vector in our notation. However, because of the
condition \eqref{eq:form:m_mu_otrh} only two components of the vector are independent. Thus, the
eigenvalue problem \eqref{eq:form:main_dynamic} is degenerate. This degeneracy can be avoided by
using circular coordinates of the Holstein-Primakoff transformation~\cite{bib:Krivosik:2010},
however, we keep three-dimensional presentation of the vector $\vec{m}_i$ to retain notational flexibility. Of course, during actual numerical simulations the non-physical solutions
should be discarded. This can be easily done, as in \eqref{eq:form:main_dynamic},
all the non-physical solutions ($\vec{m}_{i} || \vec{\mu}_i$) have zero eigen-frequencies,
$\omega = 0$.

\section{Spin wave dynamics in a periodic lattice}
\label{sec:mult}
\subsection{Multi-vector notation}
\begin{figure}[!t]
    \center\includegraphics{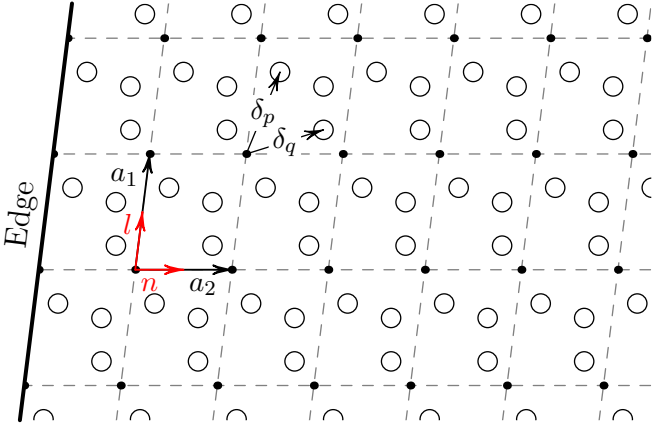}
    \caption{A sketch of a semi-infinite spatially periodic array of magnetic nanodots with a
        complex elementary cell. The primitive vectors of the lattice are $\vec{a}_1$ and
        $\vec{a}_2$, while vectors $\vec{\delta}_p$ define the positions of nanodots within the cell.
        Directions of $l$ and $n$ cell enumeration are shown in red.}
    \label{fig:complex_crystal}
\end{figure}

Both bulk~\cite{bib:Verba:2012} and edge~\cite{bib:Lisenkov:2014} collective spin wave excitations were considered previously in the arrays having a~\emph{single dot in a primitive cell}. Below, we will develop an analytical model for a periodic lattice with a non-trivial primitive cell containing several identical magnetic nanodots (having several sublattices), as shown in
Fig.~\ref{fig:complex_crystal}. For this purpose we introduce a \emph{multi-vector} notation, which will allow us to significantly simplify the theoretical formalism for the case of arrays having multiple magnetic dots in a primitive cell and, basically, to formally reduce it  to the well-developed formalism  describing dynamics in arrays with a simple primitive cell\cite{bib:Lisenkov:2014}.

The lattice of the array is defined by the primitive vectors $\vec{a}_1$ and $\vec{a}_2$, and the primitive cell consists of $P$ identical dots. Relative positions of the dots inside the primitive cell are defined by a set of vectors $\vec\delta_{p}$, $p \in [1,P]$ defined in the local coordinate system of the cell, see~Fig.~\ref{fig:complex_crystal}.  The static and dynamic magnetization of the dots in the cell are defined as $\vec{m}_{(ip)}$ and $\vec{\mu}_{(ip)}$, and the modulus of the effective static field is denoted as $B_{(ip)}$, where $i$ is the index of a cell and $p$ denotes an individual dot inside the $i$-th cell.

Using the $(ip)$-enumeration of the dots, equation \eqref{eq:form:main_dynamic} takes the following form:
\begin{equation}
	-i\omega\op{J}_{(ip)}\cdot\vec{m}_{(ip)} =
    \sum_j\sum^P_{q=1}\op{\Omega}'_{(ip)(jq)}\cdot\vec{m}_{(jq)},
	\label{eq:mult:all_index}
\end{equation}
where
\begin{equation}
\op\Omega_{(ip)(jq)} = \gamma B_{(ip)}\delta_{pq}\delta_{ij}\op I + \gamma \mu_0 M_s \op
N(\vec{r}_i-\vec{r}_j + \vec\delta_{pq}),
\label{eq:omega_general}
\end{equation}
and $\vec\delta_{pq} = \vec\delta_p - \vec\delta_q$. Unfortunately, this representation is rather cumbersome and restricts further analysis, as it obscures the structure of the problem.

To simplify equation \eqref{eq:mult:all_index} we introduce a \emph{multi-vector} of the length $P$ as:
\begin{equation}
    \mvec{m}_j = \begin{pmatrix}\vec{m}_{j1}\\\vec{m}_{j2}\\\vdots\\\vec{m}_{jP}\end{pmatrix}\label{eq:mult:multi-vector}\end{equation}
Analogously, we introduce multi-vectors for the static components of the magnetization $\mvec\mu_j = \begin{pmatrix}\vec\mu_{j1}, \vec\mu_{j2}, \cdots,
    \vec\mu_{jP}\end{pmatrix}$. In the following the multi-vectors are used as ordinary vectors in all the vector-vector and vector-matrix operations. As we will see further, the multi-vector notation is rather convenient in  both the analytical and numerical analysis. The algebra of the \emph{multi-vectors} is briefly formulated in Appendix~\ref{app:mva}.

Rewriting equation \eqref{eq:mult:all_index} by replacing the sum over $q$ with a formally written matrix-vector multiplication and by using the multi-vector notation \eqref{eq:mult:multi-vector}, one can represent it in the following form:
\begin{equation}
    -i\omega\mop{J}_i \cdot\mvec{m}_{i} = \sum_{j}\mop{\Omega}'_{ij}\cdot \mvec{m}_{j},
    \label{eq:mult:dynamic}
\end{equation}
where $i,j$ denote the \emph{cell} index, $\mop{\Omega}'_{ij}$ and $\mop{J}_i$ are the $3P\times 3P$ block-matrices defined as:
\begin{gather}
    \mop{J}_i = \diag(\op{J}_{i1},\cdots, \op{J}_{1P}),\\
    \mop{\Omega}'_{ij} =  \mop{P}_i\cdot\mop{\Omega}_{ij}\cdot\mop{P}_j,
\end{gather}
and $\mop{P}_i = - \mop{J}_i\cdot\mop{J}_i$.
Here, the tensor $\mop{\Omega}_{ij}$ is also represented by a block-matrix:
\begin{equation}
    \mop\Omega_{ij} = \gamma \delta_{ij}\mop{B}_j + \gamma \mu_0 M_s \mop{N}(\vec{r}_i-\vec{r}_j),
    \label{eq:mult:omega}
\end{equation}
where $\mop{B}_i = \diag(B_{(i1)}\op{I},\cdots, B_{(iP)}\op{I})$
and
\begin{equation}
    \mop{N}(\vec{r}) =
    \begin{pmatrix}
        \op{N}(\vec{r}) & \cdots & \op{N}(\vec{r}+\vec{\delta}_{1P}) \\
        \vdots & \ddots & \vdots\\
        \op{N}(\vec{r}+ \vec{\delta}_{P1}) & \cdots & \op{N}(\vec{r})
    \end{pmatrix}.
    \label{eq:mult:Nr}
\end{equation}
The block-matrix $\mop{\Omega}_{ij}$ describes the behavior and interactions of a~\emph{whole primitive cell} of the array in contrast to the tensor $\op{\Omega}_{ij}$ (defined by equation \eqref{eq:form:omega_general}), which describes the behavior of a~\emph{single dot}. The block-matrix $\mop\Omega'_{ij}$ is the projection of the block-matrix $\mop\Omega_{ij}$ in the same sense as presented by equation \eqref{eq:form:projection}.

Using the same procedure, one can rewrite the  equation \eqref{eq:form:sum_static} for static magnetization in terms of multi-vector notation in the following form:
\begin{equation}
    \mop{B}_i\cdot\vmu_i = \mvec{B}^\text{ext}_i - \mu_0 M_s\sum_j\mop{N}_{ij}\cdot\vmu_j,\label{eq:mult:static}
\end{equation}
where the ``components'' of the multi-vector $\mvec{B}^\text{ext}_i$ represent the external magnetic fields acting on the each dot in the $i$-th cell.

\subsection{Edge modes}

In this section we consider collective spin wave excitations in a semi-infinite array. The array
occupies a half-plane and has an edge parallel to one of the lattice primitive directions, see Fig.\ref{fig:complex_crystal}. In this case the static magnetic properties of the array demonstrate a translation symmetry along one of the principal directions, but there is no translational symmetry along the other lattice direction. Using this arrangement, we will develop an analytical model to describe the collective spin wave excitation localized near the edges of a semi-infinite array of magnetic dots.

We define the position vector $\vec r_i$ of $i$-th cell as:
\begin{equation}
    \vec r_i = \vec r_{(nl)} = l\vec{a}_1 + n\vec{a}_2,
\end{equation}
where $l$ and $n$ are the integers enumerating cells in the directions $\vec{a}_1$ and $\vec{a}_2$, respectively (Fig.~\ref{fig:complex_crystal}). The half-plane occupied by the array corresponds to $n\geqslant 0$, while the index $l$ can take any integer value.

 Since the array has a translational symmetry along the $\vec a_1$ direction, the static configuration of the magnetization depends only on the index $n$:
\begin{equation}
    \begin{gathered}
    \vmu_i = \vmu_{(nl)} = \vmu_n,\\
    \B_i = \B_{(nl)} = \B_n,
    \label{eq:mult:static_quant}
    \end{gathered}
\end{equation}
and elementary spin wave solutions $\vec m_i$ can be found in the form
\begin{equation}
    \m_i = \m_{(nl)} = \m_{n}e^{i\kappa a_1l},
    \label{eq:plane_wave}
\end{equation}
where $\kappa$ is the wave-number of the spin wave mode.
 Using these expressions in equation \eqref{eq:mult:dynamic} yields the equation for the spin wave profile $\m_n$:
\begin{equation}
    -i\omega \mop{J}_n\cdot\m_{n} = \sum_{n'=0}^{\infty}\mop\Omega'_{\kappa, nn'}\cdot \m_{n'},
    \label{eq:mult:edge_dynamic}
\end{equation}
where
\begin{equation}
    \mop\Omega_{\kappa, nn'} = \gamma  \delta_{nn'}\mop B_n + \gamma \mu_0 M_s \mop E_\kappa(n-n'),
    \label{eq:main_1D}
\end{equation}
and
\begin{equation}
    \mop E_\kappa(n) = \sum_{l}\mop{N}(l\vec{a}_1 + n\vec a_2)\cdot e^{-i\kappa a_1l}.
    \label{eq:mult:Ekreal}
\end{equation}
To compute the sum in \eqref{eq:mult:Ekreal} we introduce the Fourier transform of the cell mutual demagnetizing tensor:
\begin{equation}
    \mop{N}(\vec{r})=\dfrac{1}{(2\pi)^2}\iint\mop{N}_\vec{k}e^{i\vec{k}\cdot\vec{r}}d\vec{k}.
\end{equation}
Using the well-known ``shift property'' of the Fourier transform we obtain from \eqref{eq:mult:Nr}:
\begin{equation}
    \mop{N}_\vec{k} =
    \begin{pmatrix}
        \op{N}_\vec{k} &\cdots& \op{N}_\vec{k}e^{-i\vec{k}\cdot\vec{\delta}_{1P}} \\
        \vdots & \ddots & \vdots\\
        \op{N}_\vec{k}e^{i\vec{k}\cdot\vec{\delta}_{1P}} &\cdots &\op{N}_\vec{k}
    \end{pmatrix}.
    \label{eq:mult:Nk}
\end{equation}
Considering that $\op{N}_\vec{k}$ is a real and symmetric tensor~\cite{bib:Beleggia:2004} in the
sense that $\op{N}_\vec{k}=\op{N}^T_\vec{k}=\op{N}_{-\vec{k}}$, the Fourier image of the cell
demagnetization block-matrix has the following symmetry properties:
\begin{equation}
    \mop{N}_\vec{k} = \mop{N}^*_\vec{k}=\mop{N}^T_{-\vec{k}}.
    \label{eq:mult:Nk_prop}
\end{equation}

Using the expression for the  Fourier image \eqref{eq:mult:Nk} for $\mop{N}(\vec{r})$  equation \eqref{eq:mult:Ekreal} one gets:
\begin{widetext}
\begin{equation}
    \begin{aligned}
    \mop{E}_{\kappa}(n) =
    \dfrac{1}{(2\pi)^2S}&\sum_{l}\iint\mop{N}_{\alpha\vec{K}_1+\beta\vec{K}_2}
        e^{i(\alpha\vec{K}_1+\beta\vec{K}_2)\cdot(l\vec{a}_1+n\vec{a}_2)}e^{-i\kappa
            a_1l}\mathrm{d}\alpha\mathrm{d}\beta,
    \end{aligned}
    \label{eq:mult:Ek_obscure}
\end{equation}
\end{widetext}
where $\vec{K}_1$ and $\vec{K}_2$ are the reciprocal primitive vectors of the lattice and $S$ is the area of a unit cell. Using the
well-known properties of the reciprocal primitive vectors and the identity $\sum_le^{i\alpha l} \equiv2\pi\sum_l\delta(2\pi l-\alpha)$, equation \eqref{eq:mult:Ek_obscure} can be simplified:
\begin{equation}
    \mop{E}_{\kappa}(n)  = \dfrac{1}{2\pi
          S}\sum_{l}\int_{-\infty}^\infty\mop{N}_{(l+{\kappa a_1}/{2\pi})\vec{K}_1+\beta\vec{K}_2}e^{i2\pi
          n\beta}\mathrm{d}\beta.
    \label{eq:mult:Ek}
\end{equation}

The analytical expressions for the Fourier image of the dot demagnetization tensor $\mop{N}_\vec{k}$ are known for all the practically interesting dot shapes (see equations~(2.5)--(2.8) in [\onlinecite{bib:Verba:2012}]). Therefore, equation \eqref{eq:mult:Ek} allows one to calculate the components of the block-matrix $\mop{E}_\kappa(n)$ that are necessary for the solution of the equation \eqref{eq:mult:edge_dynamic}.

In general, the block-matrix $\mop E_{\kappa}(n)$ is not Hermitian and has no
symmetry in respect to Hermitian conjugation. However, from the properties of the
Fourier transform and the properties of the block-matrix $\mop{N}_\vec{k}$ in the equation \eqref{eq:mult:Nk_prop} one can conclude that it has the following symmetry properties:
\begin{equation}
    \mop E_\kappa(n) = \mop E^*_\kappa(-n) = \mop E^T_{-\kappa}(-n) = \mop E^*_{-\kappa}(n),
\end{equation}
and
\begin{equation}
    \Tr\left(\mop E_\kappa(n)\right) = \delta_{n0}\left(P - \sum_{p=1}^P \dfrac{K^a_p}{\muz\Ms^2}\right).
\end{equation}

The static equation~\eqref{eq:mult:static} can be rewritten in a similar way, considering that
 the static quantities \eqref{eq:mult:static_quant} do not change in the direction of the vector $\vec{a}_1$:
\begin{equation}
        \B_i\cdot\vmu_n = \mvec{B}^\text{ext}_n -
        \muz\Ms\sum_{n'=0}^{\infty}\mop{E}_0(n-n')\cdot\vmu_{n'}.
        \label{eq:mult:edge_static}
\end{equation}

Together, equations \eqref{eq:mult:edge_dynamic} and \eqref{eq:mult:edge_static} are the central result of this paper. Using these equations one can calculate the distribution of the internal magnetic field~($\B_i$)in the array, equilibrium directions of the magnetic moments in each dot~($\vmu_i$), and the spectrum of collective spin wave edge excitations in a semi-infinite array of magnetic nanodots with a complex primitive cell.

Although equations \eqref{eq:mult:edge_dynamic} and \eqref{eq:mult:edge_static} were obtained using purely analytical methods, further analytical solution of these equations (e.g., using the nearest neighbor approximation~\cite{bib:Bondarenko:2014}) is tedious and does not allow one to obtain qualitative analytical results in a closed form. The internal structure of the
tensors~$\mop E_\kappa(n)$ is complicated, and, moreover, the dispersion equation
\eqref{eq:mult:edge_dynamic} leads to a generalized eigenvalue problem with an infinite block
``Toeplitz-like'' matrix, in which blocks standing on the main diagonal are not equal, because $\mop B_n \neq \mop B_m$ when $n\neq m$ in \eqref{eq:main_1D}.

Certain analytical results may be obtained in a rather artificial case when it is assumed that the internal magnetic field $\B_n=\B$ is uniform (in this case $\mop\Omega_{nn'} = \mop\Omega_{n-n'}$, which simplifies the solution of \eqref{eq:mult:edge_dynamic}
considerably). However, even in this case the analytical calculation of the eigenvalues of the equation \eqref{eq:mult:edge_dynamic} with a non-Hermitian block Toeplitz matrix is non-trivial
\cite{bib:Tilli:1998}, and the spectrum of collective spin wave excitations of the array may include distinct edge modes~\cite{bib:Lisenkov:2014}.

Therefore, in this paper we solve~\eqref{eq:mult:edge_dynamic} and~\eqref{eq:mult:edge_static}
numerically. The details of the numerical procedure are discussed in Sec.\ref{sec:examples:pecu}. We would like to stress, that even for the numerical solution of the equation \eqref{eq:mult:edge_dynamic}, the knowledge of the analytical structure of this equation provides significant advantages. For example, one can scale the results obtained numerically for relatively small systems for the case of much larger systems and can calculate spin wave damping and FMR absorption spectra without additional numerical analysis (see Sec.~\ref{sec:fmr} for further details).

Equations \eqref{eq:mult:edge_dynamic} and~\eqref{eq:mult:edge_static} were derived for a semi-infinite array of magnetic nanodots, but this specific geometry is only reflected in the range of summation over the index $n'$ ($n\geqslant 0$, i.e., half-plane). The same equations with different summation ranges describe a number of other geometries, that are characterized by a translational symmetry along only one direction ($l$). In particular, equations \eqref{eq:mult:edge_dynamic} and~\eqref{eq:mult:edge_static} with unrestricted $n'$ should describe spin waves in an infinite array (see the next section). These equations, without any further modifications, can be used to describe the spin waves in a finite ``stripe'' ($0\leqslant n< \mathfrak{N}$) of a magnetic dot array, or the waves propagating along the internal
``domain walls'' in an infinite array. The latter example is considered in more details in
Sec.~\ref{sec:examples:dw}. In  all the following sections we shall not indicate explicitly the range of summation over the transversal indices ($n$, $n'$), assuming that they take physically relevant values in each particular case.

\subsection{Bulk modes}
In the previous section, we developed a model describing the dynamics of a magnetic dot array under the assumption that the array has a periodic translational symmetry along the vector $\vec{a}_1$, but that the translational symmetry is broken along the vector $\vec{a}_2$ due to the existence of the array's boundary. Nevertheless, the model should be valid for the cells located far from the boundary, where the edge effects are negligible. Here, we show, that for an infinite array (or for the cells located far from the boundary) our theory can be reduced to the previously developed theory of infinite arrays~\cite{bib:Verba:2012}. Also we provide a link between the tensor $\mop{E}_\kappa(n)$ and the fundamental tensor $\op{F}_\vec{k}$ introduced earlier.

To consider the case of an infinite dot array we assume that the translational symmetry also holds in the direction parallel to the primitive vector $\vec{a}_2$, meaning that all the cells in the array are equal: $\vmu_n = \vmu$, and $\B_n = \B$ for all $n$. In this case we can introduce a wave-number $\kappa_\perp$ describing the wave profile in the $\vec{a}_2$ direction. Now, the solution can be found in the form:

\begin{equation}
    \m_n = \m e^{i\pkappa a_2 n}.
    \label{eq:mult:plane_wave_perp}
\end{equation}
Direct substitution of~\eqref{eq:mult:plane_wave_perp} into the dynamic equation \eqref{eq:mult:edge_dynamic} yields a simple equation:
\begin{equation}
    -i\omega\J\cdot \mvec{m} = \mop{\Omega}'_{\vec{k}}\cdot \mvec{m},
    \label{eq:mult:bulk}
\end{equation}
which coincides with equation (3.32) from Ref.~[\onlinecite{bib:Verba:2012}], obtained for bulk spin wave modes in an infinite array of nanodots. Here $\vec{k}$ is the total spin wave wave-vector:
\begin{equation}
    \vec{k} = \dfrac{\kappa a_1}{2\pi}\vec{K}_1 + \dfrac{\pkappa a_2}{2\pi}\vec{K}_2.
\end{equation}
The interaction matrix $\mop\Omega_\vec{k}$ has the following form:
\begin{equation}
    \mop{\Omega}_\vec{k} = \gamma \mop{B} + \gamma \mu_0 M_s
    \mop{F}_\vec{k},
\end{equation}
where the block-matrix $\mop{F}_\vec{k}$ (fundamental tensor of the array~\cite{bib:Verba:2012}) may be found as:
\begin{equation}
    \mop{F}_\vec{k} =\sum_n\mop{E}_\kappa(n)e^{-i\pkappa a_2 n} =
    \dfrac{1}{S}\sum_{\vec{K}\in\mathcal{L}^*}\mop{N}_{\vec{k}+\vec{K}},
    \label{eq:mult:FktoEk}
\end{equation}
where $\mathcal{L}^*$ represents the reciprocal lattice of the array. We can use
expression \eqref{eq:mult:FktoEk} to relate the tensors $\mop{F}_\vec{k}$ and $\mop{E}_k(n)$ as:
\begin{equation}
    \mop{E}_{\kappa}(n) = \dfrac{1}{2\pi}\int_{-\infty}^\infty\mop{F}_{\kappa a_1/(2\pi)\vec{K}_1 +
        \beta\vec{K}_2}\,e^{i2\pi n\beta}\mathrm{d}\beta.
    \label{eq:mult:EktoFk}
\end{equation}
This equation provides a way of computing the block-matrix $\mop{E}_k(n)$ from the array's
fundamental tensor $\mop{F}_\vec{k}$ by performing only a single Fourier transform.

Another consequence of \eqref{eq:mult:EktoFk} is that the solutions of the eigenvalue problem
\eqref{eq:mult:edge_dynamic} always include solutions of \eqref{eq:mult:bulk}, i.e., numerically
obtained spectrum of \eqref{eq:mult:edge_dynamic} will also contain the bulk spectrum of an infinite array.

To distinguish the localized edge spin wave modes from the set of all the other solutions of the
equation~\eqref{eq:mult:edge_dynamic}, the \emph{spin wave profile} (distribution of the spin wave amplitude $|\mvec{m}_n|$) for each mode should be analyzed. Edge modes will have a stronger magnetization near the edge, while the profiles of the bulk modes do not decay into the depth of the array.

\section{FMR excitation in confined arrays}
\label{sec:fmr}
\subsection{Stationary amplitudes of forced edge modes}
The excitation and damping of collective spin waves in arrays of magnetic nanodots may be considered in terms of a standard perturbation theory. Perturbation theory was previously used for a finite array in~[\onlinecite{bib:Verba:2012}], however, that solution has a significant drawback, namely, it requires  the information about the magnetization dynamics in each particular dot, which makes the computation in  the case of arrays containing a significant number of dots ineffective and time consuming.

Let us, first, consider a semi-infinite array. As it was shown in the previous section, a
semi-infinite array has two types of modes: bulk modes and edge modes. The edge modes are localized in the area near the edge of the array, while bulk modes exist throughout the array.  From \eqref{eq:mult:edge_dynamic}, one may easily obtain the following normalization conditions for the  bulk spin wave modes:
\begin{subequations}
    \begin{equation}
        \sum_n\mvec{m}^*_{s',\kappa, n}\cdot\mop{J}_n\cdot\mvec{m}_{s,\kappa,n} =
        -iA_{s,\kappa}\delta_{s,s'},\label{eq:fmr:norm:bulk}
    \end{equation}
and, likewise, for the edge modes:
    \begin{equation}
        \sum_n\mvec{m}^*_{\nu',\kappa, n}\cdot\mop{J}_n\cdot\mvec{m}_{\nu, \kappa,n} =
        -iA_{\nu,\kappa}\delta_{\nu,\nu'}
    \end{equation}%
    \label{eq:fmr:norm}%
\end{subequations}
Here the indices $s$ denote the bulk mode zones, while the indices $\nu$ denote the localized edge modes. The bulk and edge modes are orthogonal to each other in the sense of equation \eqref{eq:fmr:norm}.

The norm defined by  equation \eqref{eq:fmr:norm:bulk} is not practical for the numerical computations, as it cannot be obtained directly from numerical solutions of equation \eqref{eq:mult:edge_dynamic}. If we neglect the influence of the boundary on the bulk
modes~\cite{bib:Lisenkov:2014} we can eliminate the summation along the row $n$ using the
dependence \eqref{eq:mult:plane_wave_perp} and come to a more practical expression:
\begin{equation}
    \mvec{m}^*_{s',\vec{k}}\cdot\mop{J}_0\cdot\mvec{m}_{s,\vec{k}} = -iA_{s,\vec{k}}\delta_{s,s'},
    \tag{\ref{eq:fmr:norm:bulk}*}
\end{equation}
where $\mop{J}_0$ describes the directions of the magnetic moments in a primitive cell situated far from the edge. The norm $A_{s,\vec{k}}$ can be directly calculated from the numerical solution of equation \eqref{eq:mult:bulk}.

Both the damping and driving microwave fields may be included in \eqref{eq:form:LL} as
small magnetic fields $\vec{b}_i(t)$ acting on each dot in the array:
\begin{equation}
    \dfrac{d\vec{M}_i}{dt} = \gamma(\vec{B}_i\cross\vec{M}_i) + \gamma(\vec{b}_i\cross\vec{M}_i).
    \label{eq:fmr:LL_with_field}
\end{equation}

First of all, we are interested in the excitation of spin wave modes of the array by a \emph{spatially uniform} external magnetic field ($\mvec{b}_0$).  Thus the perturbative magnetic field $\mvec{b}_i$ can be written as:
\begin{equation}
    \mvec{b}_i = -\dfrac{\alpha_G}{\gamma}\dfrac{d}{dt}\mvec{m}_i + (\mvec{b}_0 e^{-i\omega_0 t} \cc),
\end{equation}
where $\alpha_G$ is the Gilbert damping constant, $\omega_0$ is the excitation frequency, and c.c. denotes a complex-conjugate.

In the case of a spatially uniform excitation only the edge modes with $\kappa=0$ and the bulk modes with $\vec{k}=0$ are effectively excited. This allows us to consider the solutions
for \eqref{eq:fmr:LL_with_field} only for a single row of cells, as in Fig.~\ref{fig:complex_crystal}. The solution for the perturbed equation \eqref{eq:fmr:LL_with_field} may be written as:
\begin{equation}
    \begin{aligned}
    \mvec{m}_n(t) = &\sum_\nu c_\nu\mvec{m}_{\nu, n}e^{-i\omega_0 t} +\sum_s
    c_s\mvec{m}_{s}e^{-i\omega_0 t} \cc,
    \end{aligned}
    \label{eq:fmr:decomposition}
\end{equation}
where $c_\nu$ and $c_s$ are the stationary amplitudes of edge and bulk modes, $\mvec{m}_{\nu,n}$ are the localized solutions of \eqref{eq:mult:edge_dynamic}, and $\mvec{m}_s$ are the solutions
of \eqref{eq:mult:bulk}. To find the amplitudes of the excited modes, one should plug \eqref{eq:fmr:decomposition} into \eqref{eq:fmr:LL_with_field} and use the normalization conditions \eqref{eq:fmr:norm} to separate the equation of motion into a set of equations for individual spin wave amplitudes. In the first approximation we can neglect all the terms that are non-linear in $\mvec{m}_n$, and for the non-degenerate modes the stationary amplitude of each of the edge modes is given by
\begin{equation}
    c_\nu = \dfrac{\gamma\beta_\nu}{\delta\omega_\nu - i\Gamma_\nu},
    \label{eq:fmr:cnu}
\end{equation}
where $\delta\omega_\nu = \omega_0-\omega_\nu$,
\begin{equation}
    \begin{aligned}
        \Gamma_\nu &= \dfrac{\alpha_G\omega_\nu}{A_\nu}\sum_n\m^*_{\nu,n}\cdot \m_{\nu, n},\\
        \beta_\nu &= \dfrac{1}{A_\nu}\sum_n\m^*_{\nu,n}\cdot \mvec{b}_0.
    \end{aligned}
\end{equation}
For bulk modes the stationary amplitude is defined by the expression that is similar to~\eqref{eq:fmr:cnu}, but with
\begin{equation}
    \begin{aligned}
        \Gamma_s &= \dfrac{\alpha_G\omega_s}{A_s}\m^*_{s}\cdot \m_{s},\\
        \beta_s &= \dfrac{1}{A_s}\m^*_{s}\cdot \mvec{b}_0.
    \end{aligned}
\end{equation}
Here $\Gamma_{\nu/s}$ are the Gilbert damping rates for the corresponding spin wave modes, and $\beta_{\nu/s}$ represents the coupling coefficient between the spin wave mode and the spatially uniform external driving magnetic field.

\begin{figure}
    \center\includegraphics{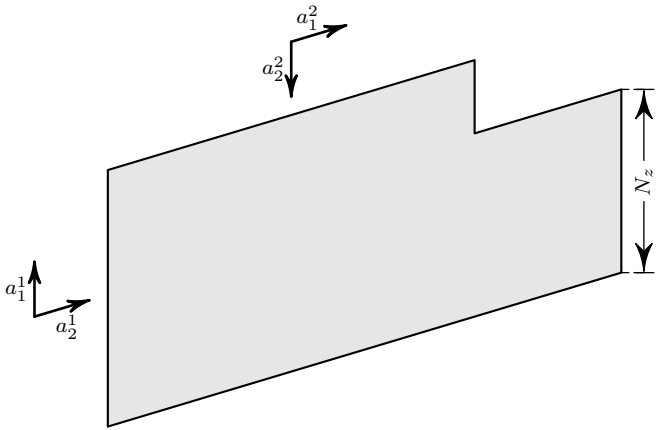}%
\caption{A sketch of a finite array in the shape of a polygon. Number of cells along each side of the polygon is $N_z$. Local sets of primitive  vectors for different sides are shown.}
\label{fig:polygon}
\end{figure}

\subsection{Power absorption of a finite polygon-shaped array}

Now let us consider a finite array containing $N$ magnetic dots in the shape of a polygon. The
polygon, shown in Fig.~\ref{fig:polygon}, has $Z$ sides and each side ($z$) contains $N_z$ dots. The array is sufficiently large, so the edge modes traveling along the different faces may be treated as independent, and the bulk modes are unaffected by the edges of the array.

The total power absorbed by a finite array can be calculated as~\cite{bib:Gurevich:1996}:
\begin{equation}
    \mathcal P = {\Ms V\omega} \sum_i\sum_{p=1}^P\Imag\left(\vec{b}^*_0\cdot
        \vec{m}_{(i,p)}\right),
    \label{eq:fmr:total_absorb}
\end{equation}
where $V$ is a volume of a single dot.

Since the magnetization amplitude of the bulk modes is identical in each unit cell of the array, it can be found as:
\begin{equation}
    \vec{m}_{p} = \sum_{s} \sum_{q=1}^P
    \dfrac{\gamma\left(\vec{m}^*_{s,q}\cdot\vec{b}_0\right)\vec{m}_{s,p}}{A_s(\delta\omega_s -
        i\Gamma_s)}.
    \label{eq:fmr:full_magn}
\end{equation}
Thus, the power absorbed (per unit cell) by the bulk modes is:
\begin{equation}
    \mathcal P_b = \dfrac{V \omega}{\muz}
    \Imag(\vec{b}^*_0\cdot\op\chi\cdot\vec{b}_0),
    \label{eq:fmr:P_b}
\end{equation}
where
\begin{equation}
    \op\chi = \sum_s\sum_{p,q}^{P}\dfrac{\omega_M}{\delta\omega_s-i\Gamma_s}
    \dfrac{\vec{m}_{s,p}\otimes\vec{m}_{s,q}}{A_s},
    \label{eq:fmr:chi_s}
\end{equation}
and $\omega_M = \gamma\muz\Ms$.

The magnetization profiles of the localized edge modes are not uniform in a direction along which the translational symmetry is lacking. To calculate the absorption caused by the edge modes localized at each edge of the array, we choose one of the local primitive vectors ($\vec{a}^z_1$) to be parallel to the chosen edge, as in Fig.~\ref{fig:polygon}. In this case we can calculate the power absorbed by a single row of primitive cells in the direction of the primitive vector $\vec{a}^z_2$:
\begin{equation}
    \mathcal P_z =\dfrac{\omega V}{\muz}
    \Imag(\vec{b}^*_0\cdot\op\chi_z\cdot\vec{b}_0),
    \label{eq:fmr:P_nuz}
\end{equation}
where
\begin{equation}
    \op\chi_z = \sum_{\nu_z}\dfrac{\omega_M}{\delta\omega_{\nu_z}-i\Gamma_{\nu_z}}\sum_{n,n'}
    \sum_{p,q}^{P}\dfrac{\vec{m}_{\nu_z,(np)}\otimes\vec{m}_{\nu_z,(n'q)}}{A_{\nu_z}},
    \label{eq:fmr:chi_nuz}
\end{equation}
and $\vec{m}_{\nu_z, (np)}$ are the localized solutions of \eqref{eq:mult:edge_dynamic} for the set of primitive vectors $\vec{a}_{1,2}^z$ and for $\kappa=0$.

Finally, the total absorption power of the array having number of cells $N$  may be calculated as:
\begin{equation}
    \mathcal P = N\mathcal P_b + \sum_z N_z\mathcal P_z.
    \label{eq:fmr:main_power}
\end{equation}
This equation demonstrates that the influence of the edge modes fades away with increasing
number of dots in the array as $N_z/N \sim 1/\sqrt{N}$.

The tensors $\op{\chi}$ and $\op{\chi}_{z}$ defined by equations \eqref{eq:fmr:chi_s}
and~\eqref{eq:fmr:chi_nuz}, respectively, have the  form of effective partial susceptibility tensors for the bulk and edge modes at each particular edge. By analyzing eigenvectors of these tensors one may find certain polarizations of external microwave field that are necessary for a maximum or/and minimum absorption caused by a certain particular mode (see examples presented below for more detail). The ability to analytically calculate the macroscopic polarization properties of an array is not only useful  for the engineering development of novel magnetic
meta-materials, but also can provide a way to experimentally verify this theory.

A separate note should be made about the ``corner modes'' of the array, e.g., the spin wave modes localized near the vertices of an array of a finite size. Obviously, the power absorption caused by these modes does not depend on number of dots in the array, so their influence is smaller than the influence of the edge modes, and, in most practical cases, can be neglected.

Finally, if the array has spatial dimensions that are much larger than the wavelength of the driving microwave field, the edge effects become less important.  However, if the tensor $\op\chi$ is known, one can calculate not only the absorbed microwave  power, but also the reflection of electromagnetic waves from an array with an arbitrarily complex primitive cell using the approximate electrodynamic boundary conditions derived in~[\onlinecite{bib:Lisenkov:2015}].

\section{Peculiarities of numerical solution}
\label{sec:examples:pecu}
To solve equations~\eqref{eq:mult:edge_dynamic} and~\eqref{eq:mult:edge_static} numerically one
should truncate the number of terms in the summation to some integer~$\mathfrak{N}$. Such a reduction of the initial problem for a semi-infinite array is equivalent to the problem of collective spin wave propagation in a \emph{finite} stripe of magnetic nanodots.  If the finite number $\mathfrak{N}$ of rows in a strip is sufficiently large, the edge spin wave modes localized at the opposite sides of the finite stripe will not interact with each other. Also, these edge modes may be degenerate (i.e. will have the identical values of the eigen-frequency ($\omega_\nu$) for certain value of wave-number ($\kappa$)). For arrays of dots having a perpendicular magnetic anisotropy the degeneracy occurs in the symmetry points of the Brillouin zone~\cite{bib:Lisenkov:2014}. In such a situation an eigenvalue solver returns vectors of the null space, which may be not localized near the opposite edges of the finite stripe, but the symmetric/antisymmetric combinations of such eigenmodes. In this case to analyze the ``real'' eigenmodes one should  construct a linear combination of those combined modes which are localized at the desired edge.

The numerical solution of the truncated eigenvalue problem \eqref{eq:mult:edge_dynamic} gives a
discrete set of eigen-frequencies and eigen-modes. To determine if a particular mode is localized near an edge, the profile of a given mode should be investigated. The numerical solution, however, cannot determine the number of localized modes, as when the truncation number $\mathfrak{N}$ increases, new localized modes may appear~\cite{bib:Landau:1977, bib:Ivanov:2014}.  However, since the main mechanism of the localization is the non-uniformity of an internal magnetic field~\cite{bib:Jorzick:2002,bib:Lisenkov:2013, bib:Lisenkov:2014} and since the magnetic field decreases as $1/\sqrt{n}$ for the dots at the position $n$ away from the array edge, these new localized modes merge with the bulk spectrum and have no influence on magnetization dynamics of the array.

Equations~\eqref{eq:mult:edge_dynamic} and~\eqref{eq:mult:edge_static} were written assuming that translational symmetry only holds along the direction of the primitive vector $\vec{a}_1$, while
such a symmetry is lacking in the direction of $\vec{a}_2$. This absence of the translational symmetry may manifest itself not only at a \emph{ physical edge} of an array, but, also, at an \emph{internal boundary} within the array, e.g., at a  domain wall or  at a peculiarity of the static magnetic field.
    For example, consider an internal boundary created by a domain wall with the following ground state distribution:
\begin{equation}
    \vmu_n = \begin{cases}\vmu_1,& \,\,\,\,\,n>0,\\ \vmu_2,& \,\,\,\,\, n\leq 0,\end{cases}
    \label{eq:examples:mmu_n}
\end{equation}
where $\vmu_1$ and $\vmu_2$ are the multi-vectors describing the stable states in the two domains separated by a domain wall. Using \eqref{eq:examples:mmu_n} in \eqref{eq:mult:edge_dynamic} and~\eqref{eq:mult:edge_static} it is possible to find a dispersion relation for the
collective spin wave modes traveling along a domain wall in the array. Thus, using the above presented theory, one can calculate all the edge, domain wall, and bulk modes in a magnetic nanodot array.  The analysis of the spatial profile of each of these modes will allow one to correctly distinguish bulk and domain-wall spin wave modes.

Truncating the summation range to $|n|<\mathfrak{N}$,  one could calculate the dispersion of spin wave modes in a \emph{stripe} of dots containing a domain wall, but, it should be noted, that using such an approximation one may also introduce some ``artifact'' modes localized near the edges of the finite-size ``stripe''.  Taking the number $\mathfrak{N}$ to be sufficiently large ensures
that the ``artifact'' modes do not interact with the domain wall modes, and by analyzing the spatial profiles of the different modes one can filter out all the ``artifact'' modes.

Of all the procedures required to obtain the numerical solution from the above presented theory, calculation of the numerical values of the block-tensor $\mop E_\kappa(n)$ requires the most processing power, since the dipolar sum in \eqref{eq:mult:Ek} converges slowly. However, the convergence rate is different for different values of $n$. Starting from a certain value $n_0$ only the elements of the sum with $l=0$ will be really significant. The exact value of $n_0$ depends on the dot shape and the interdot distance.

To compute the integral in \eqref{eq:mult:Ek}, one may also use a fast Fourier transform (FFT)
procedure or direct numerical integration, especially if only a few members of the sum are needed. Combining these techniques one can tune one's computation procedure and optimize the  time of computation.

The numerical values of the elements of the block-matrix $\mop E_\kappa(n)$ depend \emph{only} on the geometrical characteristics of the array (dot shape and lattice symmetry), and does not depend on the orientation of the  static magnetization or on the direction of the external magnetic field. Thus, once computed, the value of $\mop E_\kappa(n)$ may be saved, and, then, reused to calculate the array's characteristics for different stable states and crystalline anisotropy values, drastically reducing computation time.

\section{Examples}
\label{sec:examples}
Below we consider several examples demonstrating applications of the above presented theory for different physical systems. We show that the developed theory can be used for the calculation of the collective spin wave spectra in rather complicated systems, which may be interesting for practical applications.

In all the following examples, we will consider arrays of dots that are have a shape of round cylinders with the radius $R$ and height $h$. The static magnetic moments of all the dots are
oriented perpendicular to the array plane. If the static magnetic moments of all the dots are oriented in the same direction (either $+\vec{e}_z$ or $-\vec{e}_z$), the static state will be called \emph{ferromagnetic} (FM). If the magnetic moments of the nearest dots point in the opposite directions, the static magnetization state will be called \emph{antiferromagntic} (AFM).

A computer program implementing the proposed theoretical formalism is available
on the Internet~\cite{bib:Lisenkov:program}.

\subsection{Edge modes}
\subsubsection{FMR excitation of a finite triangular array of magnetic nanodots in
the FM static state}
\label{sec:examples:fmr}

\begin{figure}
    \center\includegraphics{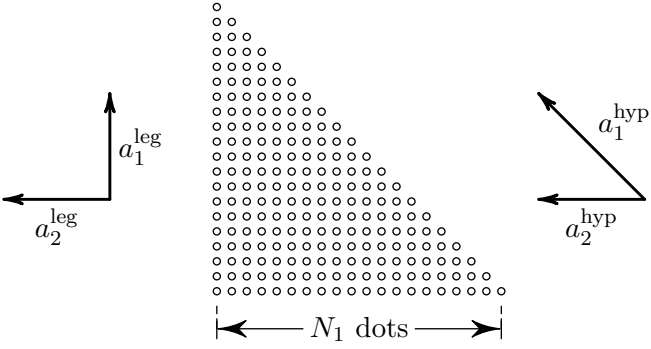}%
    \caption{Sketch of the array of magnetic nanodots in the form of a right triangle. The primitive vectors of the array's lattice used to calculate spin waves localized near hypotenuse (legs) are shown to the right (left). \label{fig:triangle}}
\end{figure}

In this example we consider an FMR excitation of a finite array of magnetic nanodots. The dots are periodically arranged in the form of a right triangle, and have a square lattice with the lattice constant $a$, as shown in Fig.~\ref{fig:triangle}. The array hs the following parameters: $a=2.2R$, $h=0.25R$, $K^a=0.5\muz\Ms^2$, $\alpha_G=0.01$, the number of dots along a leg of the
triangle is $N_1=40$, for a total of $N=820$ dots in the array. These particular parameters were chosen to guarantee the perpendicular stability of the FM state ($B^a$) and to ensure a significant dipolar interaction between the dots ($a-2R<h$).

The array considered in this example has the shape of a right triangle with three edges, two of which (the legs of the triangle) are equivalent and one is different (the hypotenuse). To employ the technique presented in Sec.~\ref{sec:fmr} we need to calculate collective spin wave spectra for two sets of the primitive lattice vectors:
\begin{subequations}
\begin{equation}
    \vec{a}^\mathrm{leg}_1 = (a,0),\,\,\,\, \vec{a}^\mathrm{leg}_2 = (0,a),
    \label{eq:examples:fmr:leg}
\end{equation}
for the modes localized near the legs and
\begin{equation}
    \vec{a}^\mathrm{hyp}_1=(a,a),\,\,\,\,  \vec{a}^\mathrm{hyp}_2 = (0,a),
    \label{eq:examples:fmr:hyp}
\end{equation}
for the modes localized near the hypotenuse.
\end{subequations}%

\begin{figure}
    \center\includegraphics{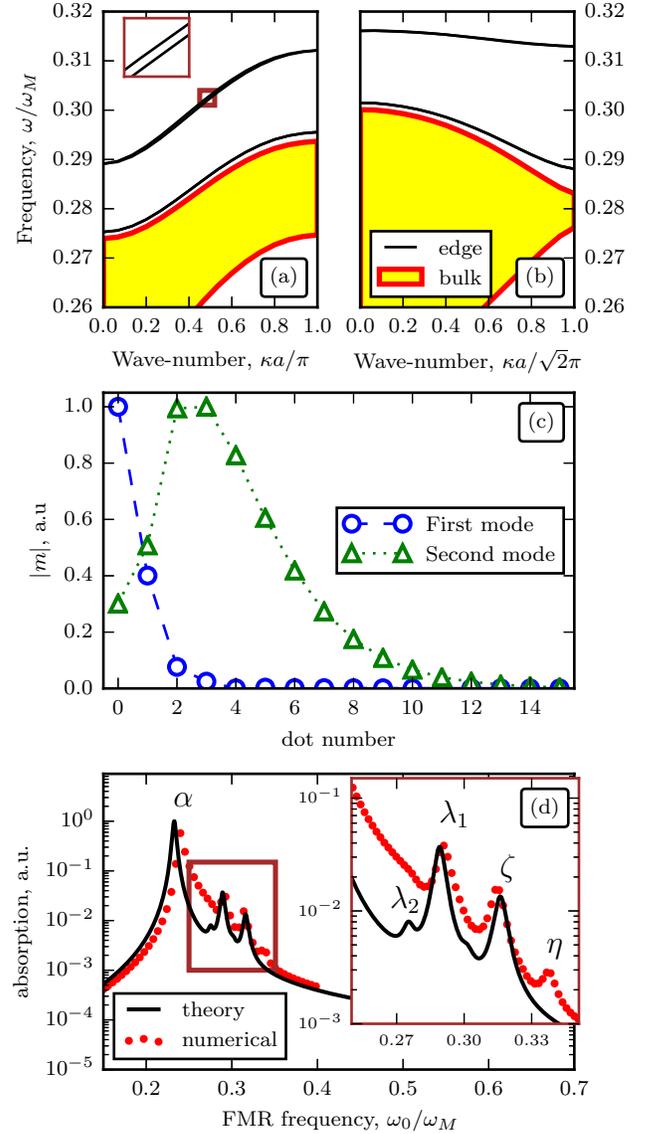}%
    \caption{Results of FMR calculation in a triangular array of magnetic dots; (a-b)~Dispersion of the spin wave modes traveling along (a)~legs and (b)~hypotenuse of the triangle. Solid lines correspond to the edge modes, yellow area denotes the bulk mode zone;  (c)~Distribution of the magnetization amplitude for two edge modes localized near the legs of the triangle;
    (d)~Absorption power spectrum of the array for a microwave signal with right circular
    polarization. The inset shows absorption caused by the edge modes in more details. The black
    line represents the calculation results based on the theory presented in this paper, while the red dots correspond to the direct numerical simulations. The absorption peaks associated with different modes are marked as follows: $\alpha$ --- bulk modes; $\lambda_1$ and $\lambda_2$ --- edge modes localized near the legs; $\zeta$ --- edge mode localized near the hypotenuse of the triangle, $\eta$ --- ``corner mode'' localized near vertexes of the triangle. Parameters of the array: $a=2.2R$, $h=0.25R$, $B^a=2.0\muz\Ms$, $\alpha_G=0.01$, the total number of dots in the array is $N=820$. The frequency values are normalized by the characteristic frequency $\omega_M = \gamma\mu_0M_s$.\label{fig:fmr}}
\end{figure}

\begin{figure}
    \center\includegraphics{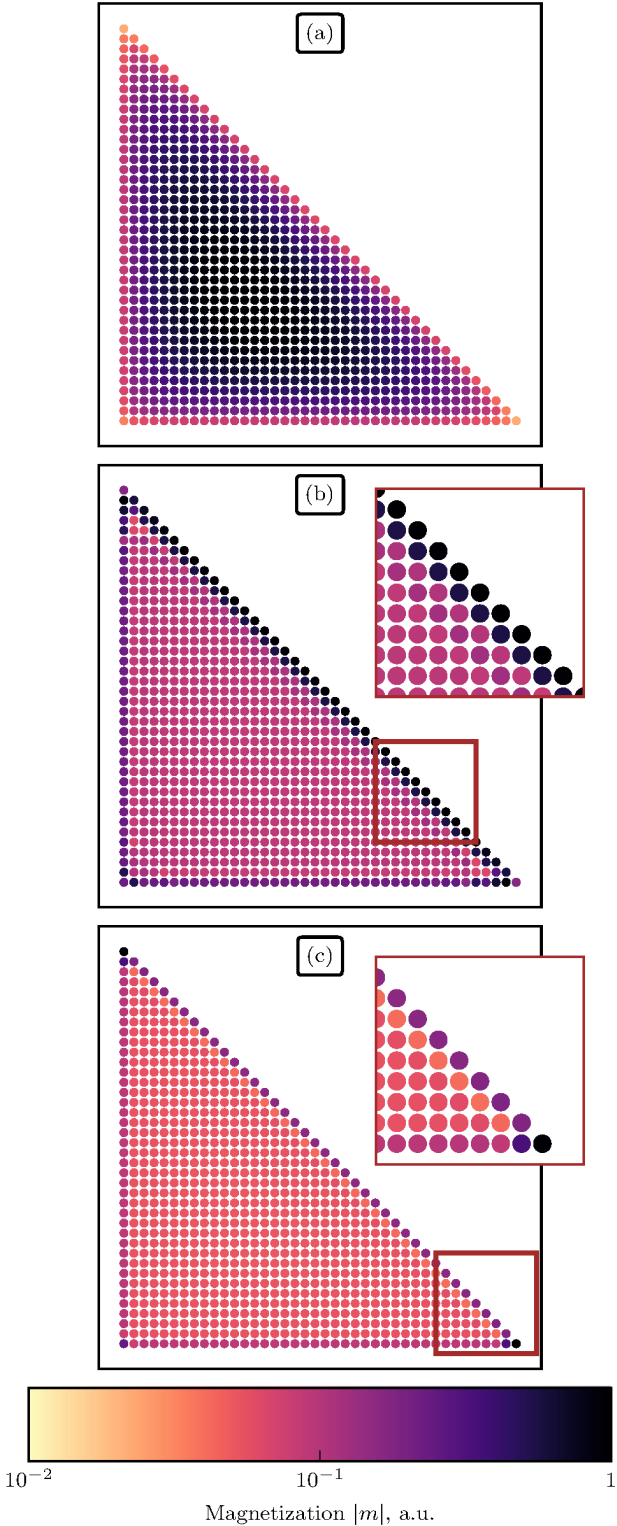}%
    \caption{Distribution of the high-frequency component of the magnetization in the triangular array of magnetic nanodots (see Fig.~\ref{fig:triangle}) excited by a variable external magnetic field, obtained by a numerical solution of the linearized Landau-Lifshitz equations. The frequency of the excitation is chosen to be at the central frequency of the absorption peaks marked in Fig.~\ref{fig:fmr}(d): (a) $\alpha$-line (bulk mode), (b) $\zeta$-line (one of the edge modes), and (c) $\eta$-line (corner mode).
     \label{fig:fmr_distrib}
    }
\end{figure}

The dispersion relations for the legs of the triangle are shown in Fig.~\ref{fig:fmr}(a), while the dispersion for the hypotenuse is shown in Fig.~\ref{fig:fmr}(b). These relations were calculated numerically using equation \eqref{eq:mult:edge_dynamic} for  equivalent stripes
oriented along the primitive vectors defined by \eqref{eq:examples:fmr:leg} and \eqref{eq:examples:fmr:hyp} with $\mathfrak{N}=31$. The properties of the spin wave modes in an FM stripe of magnetic elements was previously investigated in detail in~\cite{bib:Lisenkov:2014}. For each stripe, it is important to note, that the FMR excitations can be localized along the stripe edges, and that these edge modes form pairs, localized on the opposite sides of the stripe. The edge modes in each pair are slightly split in frequency (see the inset in~Fig.~\ref{fig:fmr}(a)), due to the symmetry breaking part of the dipole-dipole interaction~\cite{bib:Lisenkov:2014}, but the frequency splitting vanishes
at the symmetry points of the Brillouin zone. Thus, to calculate the partial permeability tensors \eqref{eq:fmr:chi_nuz} one should take a linear combination of the eigenvectors of
\eqref{eq:mult:edge_dynamic} in such a way that the high frequency components of the magnetization appear to be non-zero for dots located near {\em one side} of the stripe.  For this particular configuration two well-separated edge modes are formed for the both sets of primitive vectors. The distribution of the high frequency components is plotted in~Fig.~\ref{fig:fmr}(c) for the primitive vectors in \eqref{eq:examples:fmr:leg} for two edge modes with $\kappa=0$.

The FMR absorption spectrum of the array is plotted in Fig.~\ref{fig:fmr}(d) with a solid black line that corresponds to the results of the numerical solution of equation \eqref{eq:fmr:main_power}. As expected, the absorption spectrum consists of a main peak (marked with the symbol $\alpha$) corresponding to bulk spin wave modes in the array, and several side-peaks associated with localized edge
spin wave modes from the legs of the triangle (marked with $\lambda_1$ and $\lambda_2$) and from
the hypotenuse (marked with $\eta$). The higher edge modes are less localized (see Fig.~\ref{fig:fmr}(c)), therefore these modes are less pronounced in the absorption spectrum, and the peak associated with the second mode localized on the hypotenuse is completely suppressed by the neighboring peaks.

When the number of the dot in the array is not large, it is possible to solve the Landau-Lifshitz equation~\eqref{eq:fmr:LL_with_field} directly. Since we are interested only in the linear dynamics, we can linearize \eqref{eq:fmr:LL_with_field} using the same procedure as described in Sec.~\ref{sec:form}:
\begin{equation}
    \sum_j\left[i\omega(\op{J}_i - \alpha\op{I})\delta_{ij} + \op\Omega'_{ij}\right]\cdot
    \vec{m}_j = \gamma \op P_i \cdot \vec{b}_0\label{eq:fmr:full_lin_LL}.
\end{equation}
The static properties can be found from~\eqref{eq:form:sum_static}, where the external field
$\vec{B}^\text{ext}_i$ is absent in our case.

In Fig.~\ref{fig:fmr}(d) we also show (see red dotted line) the absorption spectrum found by the direct numerical solution of the linear inhomogeneous system of 1640 equations~\eqref{eq:fmr:full_lin_LL} for $N=820$ dots. It is evident, that the direct numerical simulation agrees reasonably well with our theory, as the frequencies and heights of the absorption peaks are in good quantitative agreement. The most notable difference between the theoretical and numerical simulations is the broadening of the bulk peak obtained in the direct numerical simulations. The direct numerical simulations also show an additional peak (marked by $\eta$) which does not correspond to any of the peaks obtained in the quasi-analytical theory.

To further analyze the FMR absorption spectra and explain the differences between the theoretical and numerical values of the power absorption, we present in Fig.~\ref{fig:fmr_distrib} the distribution of the microwave magnetization in the array for several values of the excitation frequency, calculated from~\eqref{eq:fmr:full_lin_LL}. In Fig.~\ref{fig:fmr_distrib}(a) we plot the distribution of the magnetization at the central frequency of the  bulk peak (marked with the an $\alpha$ in Fig.~\ref{fig:fmr} ). The distribution is not uniform across the array, especially at the vertexes. This non-uniform distribution is caused by a non-homogeneous internal magnetic field within the elements, and, in its turn, is causing inhomogeneous broadening of the main absorption peak~\cite{bib:Verba:2012} seen in Fig.~\ref{fig:fmr}(d).

The magnetization distribution for the hypotenuse mode (marked in Fig.~\ref{fig:fmr} with a $\zeta$) is plotted in~Fig.~\ref{fig:fmr_distrib}(b) for the central frequency of the absorption peak. As one can see, most of the dots having the maximum values of the precession amplitude are localized near the hypotenuse, as theoretically predicted. The peak marked with the symbol $\eta$ in Fig.~\ref{fig:fmr} is not present in the theoretical calculations. The distribution of magnetization at the central frequency of this peak is plotted in~Fig.~\ref{fig:fmr_distrib}(c). For this case one can see, that the dots with the maximum amplitudes are situated at the vertexes of the triangle. These ``corner modes'' were neglected in \eqref{eq:fmr:main_power}, and they do not appear in the results obtained using the developed theory. The direct numerical simulation shows that although the influence of these ``corner modes'' is rather small, one should not completely neglect them for an array of this rather small size.

Overall, our theory gives results which correspond closely with direct numerical simulations, and provide a convenient way to analyze the features of the absorption spectra analytically. We would like to emphasize the computational advantage of our theoretical method. Each red dot in~\ref{fig:fmr}(d) required an \emph{independent} solution of the inhomogeneous linear system of equations~\eqref{eq:fmr:full_lin_LL}, having the arithmetic complexity of $O(N^3)$, making the
direct numerical simulations impractical for large arrays of magnetic dots. Even for a system with only 820 dots, the time required to calculate the numerical absorption spectra was considerable.  On the other hand, the theoretical line in the same figure, along with the dispersion relation in figure \ref{fig:fmr}(a), can be almost instantaneously computed for an array of any size, independently of the number of dots. In other words, the arithmetic complexity of the quasi-analytical problem for the same array is of $O(1)$. Surely, this simple analytical technique cannot compete in numerical accuracy with the available micromagnetic packages~\cite{bib:Vansteenkiste:2014}, however, it can be useful for the approximate``engineering'' of the desired absorption spectra with a subsequent quantitative verification using a computationally-intensive micromagnetic simulation.

\subsubsection{Nonreciprocal spin wave edge mode}
\begin{figure}
    \center\includegraphics{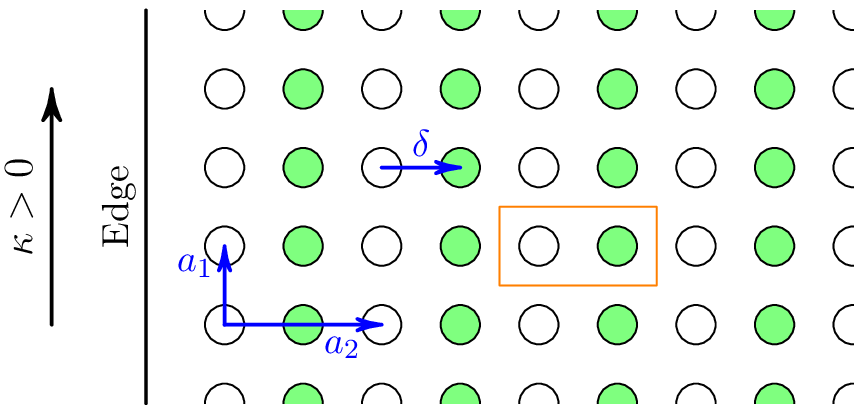}%
\caption{Sketch of a semi-infinite array of magnetic nanodots with a complex unit cell having dots with different anisotropy within a primitive cell. White and green dots have the different value of the anisotropy field $B^a$. The black arrow points towards the ``forward'' direction of the spin wave propagation. \label{fig:non_recip_scatch}}
    \center\includegraphics{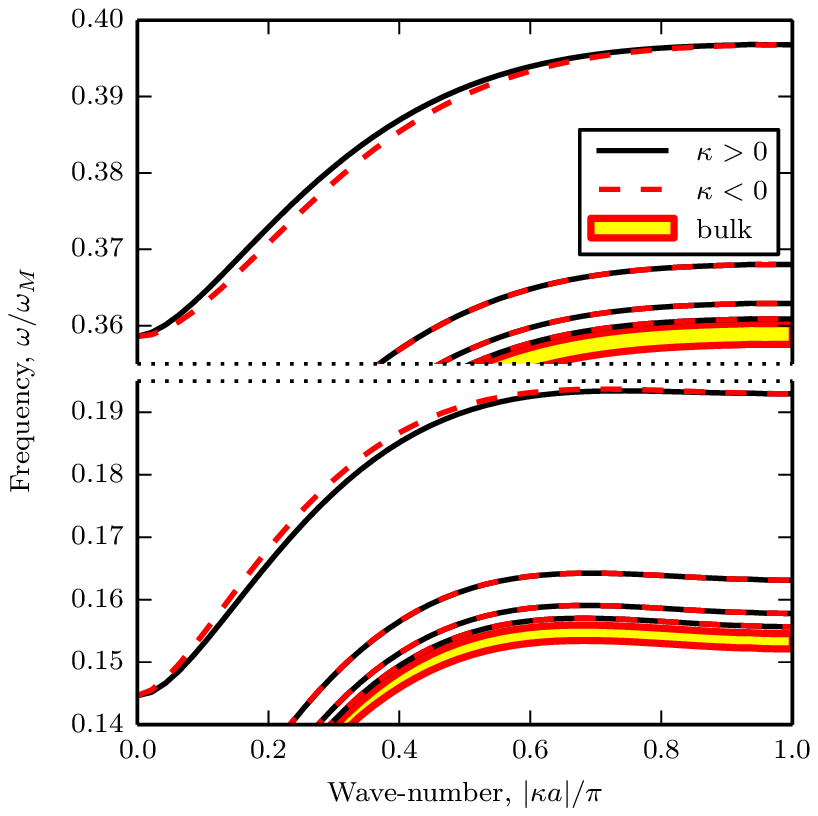}%
\caption{Dispersion of collective spin wave modes in a semi-infinite array of magnetic dots with different crystalline anisotropy, shown in~Fig.~\ref{fig:non_recip_scatch}. Dispersion curves for the edge modes traveling in the forward (backward) direction are plotted with solid (dashed) lines. The bulk spectrum is shown by a yellow region.  Parameters of the array: $a_1=3.3R$,
    $a_2 = 10R$, $\delta=5R$, $h=5.0R$, $B^a_1=0.2\muz\Ms$, $B^a_2=0$.
		The frequency values are normalized by the characteristic frequency $\omega_M = \gamma\mu_0M_s$.
    \label{fig:non_recip_disp}}
\end{figure}

This example is illustrating the non-reciprocal properties of the spin wave spectra in arrays of magnetic nanodots with complex unit cell. The non-reciprocity of a spin wave propagation is a desirable feature for the development of miniaturized and bias-free microwave isolators and circulators.  Recently, non-reciprocal bulk collective spin wave modes have been investigated
in detail for infinite arrays of magnetic nanopillars in~\cite{bib:Verba:2013}. In particular, it was shown that for arrays of identical dots, non-reciprocal spin wave bulk modes only exist when
the constituent elements have an out-of-plane magnetization.  It was also shown that non-reciprocal spin waves bulk modes exist in infinite dot arrays with complex unit cells, e.g. containing two different types of dots . For example, the dots may have different values
of the anisotropy field or different dot shapes, thus forming two different sub-lattices in the array.

In terms of the non-reciprocity the edge modes are not as restrictive as the bulk
mores,  and for the edge modes the necessary conditions for the non-reciprocal behavior could  be lifted. For example, even a simple system of identical dots may have non-reciprocal edge modes~\cite{bib:Lisenkov:2014}. The general properties of the bulk spin wave
spectra may be obtained by analyzing the symmetry properties of the fundamental tensor
$\mop{F}_\vec{k}$. For edge modes, however, an analytical solution is not possible, as the eigenvalue problem \eqref{eq:mult:edge_dynamic} is not symmetric. Nevertheless, much information can be gained by performing a numerical analysis based on \eqref{eq:mult:edge_dynamic} and
\eqref{eq:mult:edge_static}.

Here, we consider a semi-infinite array of magnetic nanodots having a rectangular lattice and a unit cell containing two different dot types, each having a different crystalline anisotropy within the same physical geometry (Fig.~\ref{fig:non_recip_scatch}). Thus, each cell of the array consists of a heterogeneous pair of dots separated by the vector $\vec\delta$. To handle this problem in the framework of our theory the diagonal components of the demagnetization block-matrix \eqref{eq:mult:Nr} should rewritten in the following way :
\begin{equation}
    \mop{N}(\vec{r}) =
    \begin{pmatrix}
        \Nform(\vec{r}) &  \Nform(\vec{r}+\vec{\delta}) \\
        \Nform(\vec{r} - \vec{\delta}) & \Nform(\vec{r})
    \end{pmatrix}
    +
    \delta(\vec{r})\begin{pmatrix}
        \op{K}_1 &  \op{0} \\
        \op{0} & \op{K}_2
    \end{pmatrix},
    \label{eq:mult:Nr_and_K}
\end{equation}
where $\op{0}$ is a zero 3x3 matrix.

For this example, when performing the actual numerical calculations of the block-matrix $\mop{E}_\kappa(n)$ using \eqref{eq:mult:Ekreal}, it is convenient to separate the part of $\mop N(\vec{r})$ that corresponds to the crystalline anisotropy. Since the anisotropy tensors $\op{K}_i$ enters directly only into the multi-tensor $\mop{E}_\kappa(0)$, it is not necessary to recalculate the numerical values of $\mop{E}_\kappa(n)$ for a different values of the anisotropy.

For the easy comparison with the previous results, we use the same geometrical parameters of the array as in~Ref.~[\onlinecite{bib:Verba:2013}]: $a_1=3.3R$, $a_2 = 10R$, $\delta=5R$, $h=5.0R$. The unit cell consists of two dots, one magnetically isotropic and another having an out of plane easy axis crystalline anisotropy of the value of $B^a_2=0.2\muz\Ms$. It has been shown previously, that if $\delta$ = $\vec{a}_2/2$, the bulk spectrum is reciprocal. However, the edge modes  in this case are not reciprocal. This is shown by the results of a numerical simulation plotted in~Fig.~\ref{fig:non_recip_disp}. The numerical simulations of edge spin waves in the array were performed for the stripe of width $\mathfrak{N}=31$ cells.

In~Fig.~\ref{fig:non_recip_disp} we present the dispersion of the edge modes traveling along one side of the stripe. The spectrum consists of two zones of bulk waves and several well-separated edge modes for each zone. The mechanism of the edge mode formation is the same as in the case considered in~Sec.~\ref{sec:examples:fmr}, namely, a non-uniform static magnetic field near the edge of the array. Corresponding to the previous results, the bulk mode spectrum is reciprocal, while the spectrum of the edge modes is non-reciprocal, because the waves traveling in the opposite directions have different eigenfrequencies. The effect of the nonreciprocity is more pronounced for the modes that are separated from the bulk spectrum. However, the difference between the frequencies for the opposite values of $\kappa$ is relatively small in this case.

The edge spin wave spectrum is reciprocal if the boundary in the array is made along the
$\vec{a}_2$ direction. The derivation of the exact analytical conditions of non-reciprocity for the spectrum of edge spin wave modes is beyond the scope of our current work  and would be published elsewhere as a separate study.

\subsection{Domain walls}
\label{sec:examples:dw}
\subsubsection{Non-reciprocal spin waves traveling along domain walls in arrays in FM stable state}
\begin{figure}
    \includegraphics{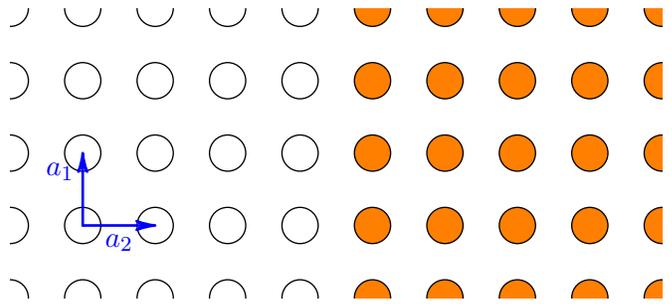}
\caption{A sketch of an array of magnetic nanodots with a domain wall in ferromagnetic static magnetization state. Open (orange) dots correspond to dots having the static magnetization directed up (down).
				}
\label{fig:domain_wall_FM_sketch}
\end{figure}

\begin{figure*}
    \center\includegraphics{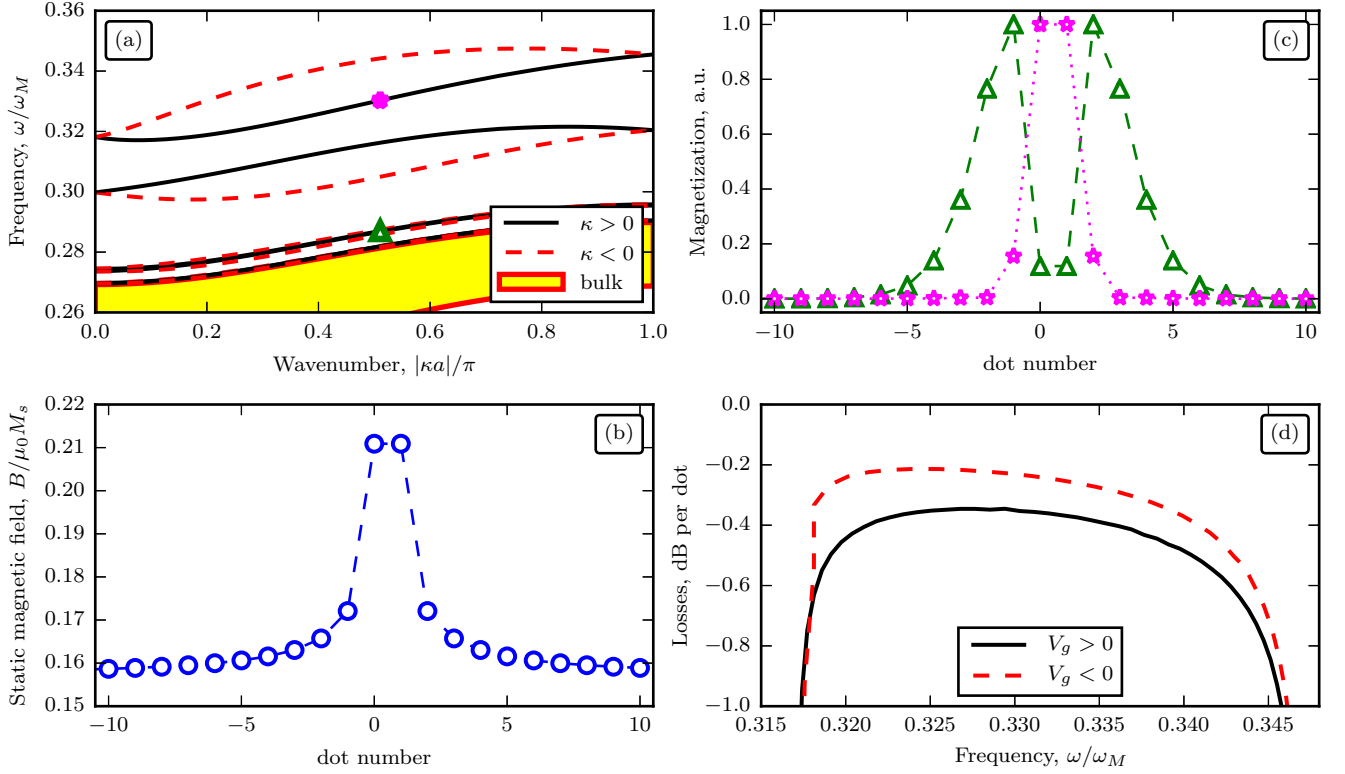}%
\caption{
    (a) Dispersion of collective spin wave excitations localized on a domain wall in an array
    of magnetic nanodots existing in the  FM static state; (b) Static magnetic field profile near a domain wall; (c) Profiles of time-averaged high-frequency component of the magnetization for the first (magenta stars) and the third (green triangles) modes traveling in the $\kappa>0$ direction. The modes are marked by similar symbols as  in the panel a);
    (d) Insertion losses for the collective spin waves traveling along the domain wall in two directions. Parameters of the array: $a=2.2R$, $h=0.25R$, $B^a=2.0\muz\Ms$,
    $\alpha_G=0.01$. The frequency values in (a) and (d) are normalized by the characteristic frequency $\omega_M = \gamma\mu_0M_s$.
		\label{fig:domain_wall_FM_disp}}

\end{figure*}

In the previous example the edge mode was well-separated in frequency from the bulk spectrum, and it exhibited a non-reciprocal behavior. While the non-reciprocity was present, the effect was relatively small. As it will be shown below, the non-reciprocal property of the spin wave modes propagating along the domain walls is substantially more pronounced than for the modes propagating along the array's edges. The reason is that the non-reciprocal frequency splitting depends on the difference in
ellipticity between the adjacent dots, and should be the most pronounced for the dots with the opposite directions of the static magnetic moments, and therefore, the opposite senses of the magnetization precession~\cite{bib:Verba:2013}.  Such a case is realized for an array in the FM stable state having a domain wall, which separates dots with anti-parallel directions of the magnetic moment, as is shown in Fig.~\ref{fig:domain_wall_FM_sketch}.

Consider an infinite array, where the static magnetization(see \eqref{eq:mult:static_quant}) has the following dependence of coordinate in the $\vec{a}_2$ direction,:
\begin{equation}
    \vec{\mu}_n = \begin{cases}-\hspace{-3 mm}&\vec{e}_z,~~~~n>0,\\ &\vec{e}_z,~~~~n\leq 0,\end{cases}
    \label{eq:examples:mu_n}
\end{equation} where $\vec{e}_z$ is a unit vector perpendicular to the array plane.

The dispersion relation for this array is shown in~Fig.~\ref{fig:domain_wall_FM_disp}(a). The numerical simulations were performed for a stripe having $\mathfrak{N}=82$ rows. This number of rows was chosen to guarantee the absence of interaction between the domain-wall modes and the ``artifact'' edge modes (the ``artifact'' edge modes have been removed from the plot).  The spectrum consists of the bulk modes, similar to the modes considered in Sec.~\ref{sec:examples:fmr}, and the domain wall modes that are separated in frequency from the bulk modes. The mechanism responsible for the formation of the domain wall modes is a ``potential well'' formed by the non-uniform static magnetic field profile near the domain wall, as shown in Fig. \ref{fig:domain_wall_FM_disp}(b). The domain-wall modes can be divided into two types: the modes localized directly on the domain wall and the modes localized near the domain wall, as shown in Fig. \ref{fig:domain_wall_FM_disp}(c). The modes localized near the domain wall (green triangles
in Fig.~\ref{fig:domain_wall_FM_disp}(c)) are similar to the modes formed near the edge of an array in FM stable state~\cite{bib:Lisenkov:2014}.  The modes formed directly on the domain wall (magenta stars in Fig.~\ref{fig:domain_wall_FM_disp}(c)) exhibit a different behavior, as these modes are more localized and are highly non-reciprocal. In fact, these non-reciprocal modes still exist in a stripe of dots consisting of only  two rows of dots with opposite directions of the static magnetization.

Non-reciprocal signal processing devices, such as isolators  and circulators, allow propagation of waves in one direction and ado not allow such propagation in the opposite direction. This property is, typically, achieved due to the different insertion losses for waves traveling in the opposite directions.

Below, we demonstrate that such a nonreciprocal isolation effect can be achieved  for the spin waves propagating along a domain wall in an array of magnetic nanodots.

The propagation losses of any of the spin wave modes caused by the magnetic damping can be calculated as:
\begin{equation}
    d_\nu = - 20 \log_{10}(e) \frac{a\Gamma_\nu}{|V^g_\nu|}\,\,\,\, [\dBperdot],
    \label{eq:examples:dBperDot}
\end{equation}
where $a$ is the distance between the nearest dots and $V^g_\nu$ is the group velocity of the spin wave mode. The direction of the wave mode propagation (``forward'' or ``backward'') is determined by the sign of the group velocity.

It is clear from Fig.~\ref{fig:domain_wall_FM_disp}(a), that there is a small region near $\kappa=0$, where the spin wave mode marked by a magenta star in Fig.~\ref{fig:domain_wall_FM_disp}(a) has a negative group velocity $V_\nu^g =d\omega_\nu/d\kappa <0$, and, therefore, can be considered as a``backward-propagating'' wave in this region.

The results of calculation of the propagation losses for spin wave modes forward- and backward-propagating along the domain wall using~\eqref{eq:examples:dBperDot} are presented in Fig.~\ref{fig:domain_wall_FM_disp}(d). In contrast with the previous example, in this case the difference in losses between the forward- and backward-propagating  wave modes is rather big (up
to \ilu[0.2]{\dBperdot}), and is comparable to the  direct insertion losses for the faster ($V_g<0$ in our case) spin wave mode. This makes the non-reciprocal spin waves propagating allong the domain walls in the FM -state magnetic dot arrays rather interesting for practical realization of nano-sized microwave isolators.

\subsubsection{Spin wave domain wall modes in chessboard AFM stable state}

\begin{figure}
    \includegraphics{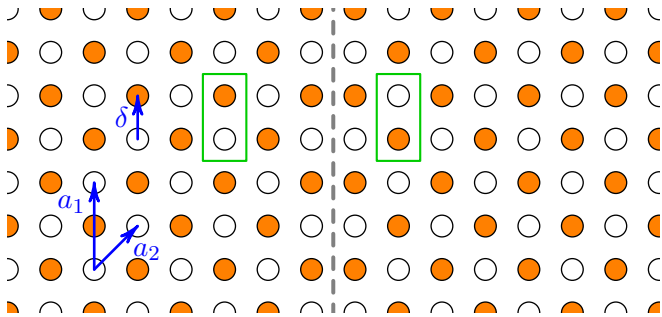}
\caption{A sketch of a domain
    wall in an array in chessboard antiferromagnetic ground state. The boundary between two states in shown by a dashed grey line. Green rectangles show the elementary
    cells for two different states. Open and orange dots has the same meaning as in
Fig.~\ref{fig:domain_wall_FM_sketch}.}
\label{fig:domain_wall_AFM_sketch}
\end{figure}

\begin{figure*}
    \center\includegraphics{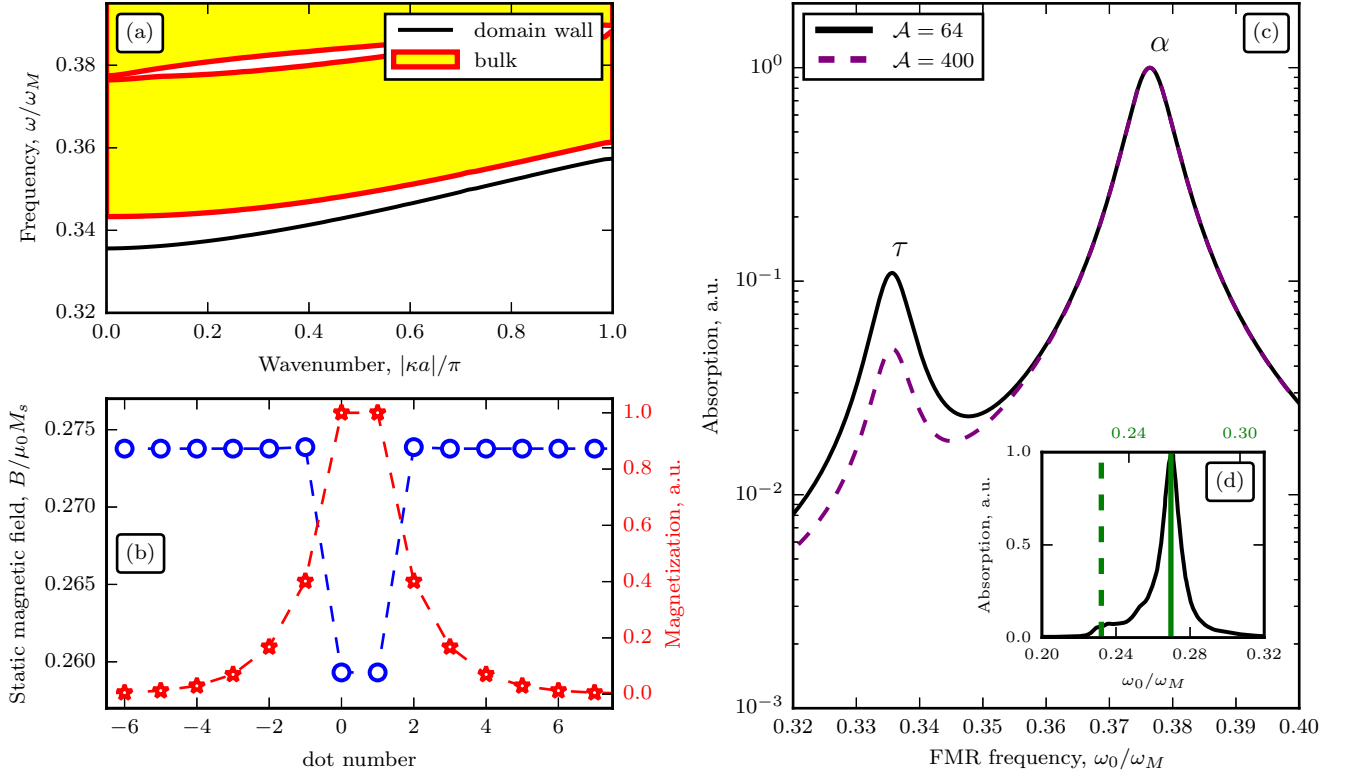}%
\caption{(a) Dispersion of the collective spin wave modes localized on a domain wall in an array
    of magnetic nanodots existing in the chessboard AFM static magnetization state; (b) Blue circles show the profile of the static internal magnetic field near the domain-wall (left axis); Red stars show the profile of the localized spin wave mode for $\kappa=0$ (right axis);(c)~Absorption power spectrum of the array in a multi-domain chessboard AFM state for the microwave signal with linear polarization oriented in the [11] direction for two values of the characteristic domain size calculated using the above developed quasi-analytical theory. The absorption peaks associated with the bulk and domain-wall modes are marked with the symbols $\alpha$ and $\tau$, respectively; (d)~Absorption power spectrum calculated using the direct micro-magnetic simulation procedure described in Ref.~\onlinecite{bib:Verba:2013a} (black line, bottom axis); Solid and dashed  vertical lines indicate theoretically calculated resonance frequencies for bulk and domain-wall modes, respectively (upper axis);
    \label{fig:domain_wall_AFM_disp} Physical dimensions of the array for (a-c) are the same
    as~for~Fig.~\ref{fig:domain_wall_FM_disp}.The frequency values in (a), (c), and (d) are normalized by the characteristic frequency $\omega_M = \gamma\mu_0M_s$.}
\end{figure*}

A magnetic dot array with a square lattice  existing in the AFM static magnetization state,
where the neighboring nanodots have opposite magnetization directions, has the lowest potential energy, and, therefore, forms a true ground state of the array. Such a state of the static magnetization, shown in Fig.~\ref{fig:domain_wall_AFM_sketch} and having a zero net magnetic moment, is, usually, naturally formed during the demagnetization and is called the chessboard AFM (CAFM) state\cite{bib:Bondarenko:2011,bib:Verba:2012}. In reality, however, demagnetization does not lead to an ideal CAFM state~\cite{bib:Verba:2013a}, but, instead, it leads to the formation of clusters with local periodicity due to the spontaneous symmetry breaking between the two equivalent ground states of the array.  A boundary between the two equivalent states CAFM states forms a ``domain wall,'' as shown in Fig.~\ref{fig:domain_wall_AFM_sketch}.

This example will explore the FMR (wave number equal to zero) absorption spectrum of the domain-wall modes in an array exising in a CAFM. To calculate the spin wave modes localized on the domain walls in the array we consider a single infinitely long domain wall within an infinite array, as shown in Fig.~\ref{fig:domain_wall_AFM_sketch}. The spectrum of the collective spin wave modes  localized on the domain wall in the CAFM state is presented in~Fig.~\ref{fig:domain_wall_AFM_disp}(a).  The spectrum consists of two bulk zones\cite{bib:Verba:2012} and a well-separated domain-wall mode. As in the previous example, we have removed the ``artifact'' modes caused by the specifics of the numerical solution (see
Sec.~\ref{sec:examples:pecu}). The main reason for the modes localization on a domain wall is
the variation of the internal magnetic field near the domain wall (see~Fig.~\ref{fig:domain_wall_AFM_disp}(b)).  However, this variation is much smaller than in the example described in the previous section: only the dots that are the closest to the domain-wall have a distinct difference in their internal magnetic field, and this field variation is rather small, making the corresponding potential well rather shallow. Therefore, only a single localized mode appears in the spectrum of the array for the considered geometrical parameters, and the localization of this mode ir rather weak as shown in Fig.~\ref{fig:domain_wall_AFM_disp}(b).

An array in the CAFM ground state does not have a net magnetic moment. Thus, its FMR absorption spectrum does not depend on the polarization of the external signal. In contrast, the contribution made by the external microwave field to the domain wall mode has a polarization dependence. If one takes the axis $Ox$ to be parallel to the domain wall, the partial susceptibility tensor \eqref{eq:fmr:chi_nuz} associated with the domain-wall modes
has the following structure:
\begin{equation}
    \op\chi_\text{dw} = \begin{pmatrix}
        \chi^{xx}_\text{dw} & 0 & 0\\
        0 & 0 & 0\\
        0 & 0 & 0
    \end{pmatrix},
\end{equation}
making the mode insensitive to the component of the external magnetic field that is perpendicular to the domain wall.

The difference in response of the  bulk and edge modes is caused by the symmetry of the problem. In a domain existing in a uniform CAFM ground state, the magnetization in a particular dot will precess clockwise and counter-clockwise depending on the orientation of the magnetic moment of this  dot~\cite{bib:Verba:2012}, making $\op\chi_\text{b}$ a diagonal real tensor with
$\chi^{xx}_\text{b}=\chi^{yy}_\text{b}$.

However, in the case of domain-wall modes the boundary between the two CAFM ground states ``synchronizes'' the precession in the adjacent domains. The mirror symmetry of the problem requires that:
\begin{subequations}
\begin{equation}
    \vec{m}_{(n+1,1)}^x=\vec{m}_{(-n,1)}^x=\vec{m}_{(n+1,2)}^x=\vec{m}_{(-n,2)}^x
\end{equation}
and
\begin{equation}
    \vec{m}_{(n+1,1)}^y=-{\vec{m}_{(-n,1)}^y} =
    -\vec{m}_{(n+1,2)}^y=\vec{m}_{(-n,2)}^y,
\end{equation}
\end{subequations}
Using the above properties on can easily to show that:
\begin{equation}
    \begin{aligned}
        \vec{m}_{(n,1)}^y\left(\vec{m}_{(n',1)}^y\right)^*=-\vec{m}_{(n,2)}^y\left(\vec{m}_{(n',2)}^y\right)^*.
    \end{aligned}
\end{equation}
 Using Eq.(65) in \eqref{eq:fmr:chi_nuz} it is easy to get that $\chi_\text{dw}^{yy} = 0$. Analogously, one can show that $\chi_\text{dw}^{xy} = 0$, and, therefore, $\chi_\text{dw}^{xx}$ is the only non-zero component of the susceptibility tensor of the domain-wall spin wave mode.

A convenient characteristic of the final state of a demagnetized array~\cite{bib:Verba:2013a} is the typical value of the number of dots forming a one CAFM cluster $\mathcal{A}$ (cluster size). Thus, the total length (in dots) of a domain wall encircling all the clusters can be approximated as:
\begin{equation}
    L \approx 2\sqrt{\mathcal{A}}\,\dfrac{N}{\mathcal{A}}.
    \label{eq:examples:domain_len}
\end{equation}
The full absorption power can be found from the  expression that is analogous to \eqref{eq:fmr:main_power}:
\begin{equation}
    \mathcal P \approx N \left( \mathcal P_b + \dfrac{2}{\sqrt\mathcal{A}} \mathcal P_\text{dw} +
        \dfrac{4}{\sqrt{N}} \mathcal P_\text{e}\right),
    \label{eq:examples:total_power}
\end{equation}
where $\mathcal P_b$ is the total absorption caused by the bulk modes, $\mathcal P_\text{dw}$  is the total absorption caused by the domain wall modes, and $\mathcal P_\text{e}$  is the total absorption caused by the edge modes.

For sufficiently large arrays the role of the edge modes vanishes, although the absorption caused by the domain wall modes is inversely proportional to the size of a cluster in the
array, and does not vanish in the limit $N\to\infty$. The absorption spectra calculated for two different values of the cluster size are shown in~Fig.~\ref{fig:domain_wall_AFM_disp}(c).  The peaks associated with the bulk modes and domain-wall modes are marked with the symbols $\alpha$ and $\tau$, respectively. For the clusters containing $\mathcal{A}=64$ dots these domain-wall modes may be clearly pronounced, producing an absorption that has a magnitude larger than 10\% of the magnitude of the bulk mode peak. Thus, the absorption caused by the domain wall mode can be easily observed experimentally. Obviously, the static magnetization states with smaller clusters produce  a larger absorption, however, when the linear dimensions of the domains get closer to the spin wave localization length (see~Fig.~\ref{fig:domain_wall_AFM_disp}(b)) the
approximation of an infinitely long domain wall is no longer valid.

A previous work \cite{bib:Verba:2013a} calculated the FMR absorption spectrum for a CAFM array
containing stable clusters using a  micro-magnetic simulation. When the data from the micro-magnetic simulation is compared with our quasi-analytical theory, the results apper to be quite similar. These results are shown in Fig. \ref{fig:domain_wall_AFM_disp}(d), where the data taken from Ref.~\onlinecite{bib:Verba:2013a} is shown by a continuous thin black line, while the vertical lines show the spectral positions of the absorption peaks calculated from our quasi-analytical. It is evident that the bulk peak obtained in the framework of our theory and shown by  a solid vertical line, matches nicely with the absorption peak obtained from the micromagnetic simulations. Likewise, the dashed vertical line, which marks the absorption peak corresponding to  the cluster of domain walls, matches well with the position of the plateau obtained in the micromagnetic simulations. Unfortunately, the data from Ref.~\onlinecite{bib:Verba:2013a} was obtained for the cluster sizes of $\mathcal{A}\approx23$, for which the localized domain wall modes were not completely formed. That makes a more detailed
quantitative comparison between the theory presented here and the micromagnetic simulations
impossible. This is confirmed  by the lack of a strong absorption peak corresponding to the domain wall modes  in the micromagnetic simulations.  The difference in the mode frequencies between the micromagnetic results and the analytical theory is caused, most likely, by the inhomogeneity of the internal magnetic field, the effect  similar to the one described in Sec.~\ref{sec:examples:fmr}.

Thus, the developed theory provides a reasonably good tool to estimate the cluster size from the FMR response measured or calculated for the arrays with sufficiently large clusters. However, for arrays with smallar clusters, the development of a technique capable of estimating the cluster size from the FMR response require a more rigorous estimation of the domains boundary length, than the one given by equation \eqref{eq:examples:domain_len}. A more rigorous estimation
must take into account the ``corner modes'' that form at the domain wall intersections and the interactions between the domain-wall modes exsisting within one domain.

\section{Conclusions}
\label{sec:conclusions}
 A general theoretical formalism, developed in this paper, allows one to calculate  the following characteristics of the finite magnetic dot arrays having a complex primitive cell and a~translational symmetry along one of the lattice vectors: (i) distribution of the internal magnetic field inside the array; (ii) equilibrium directions of the static part of the dot magnetization inside the array;  (iii) spectra of the collective spin wave edge excitations in the array. We have shown that by introducing a ``multi-vector'' notation it is possible to reduce the  solution of the Landau-Lifshitz equation for a dot array to a generalized eigenvalue problem. The components of the block-matrices involved in the eigenvalue problem for edge spin-wave excitations are obtained using a one-dimensional Fourier transform of analytically defined functions, which requires substantially less computation time than the typical micromagnetic simulations for the same system. Although only a macrospin approximation of the magnetization dynamics in a single dot is considered in this work, the developed formalism could be extended to describe the magnetization dynamics in the  dot array with non-trivial distribution of the dynamic magnetization inside the dots (for one of possible extensions see Appendix~\ref{app:ext}).

  The above developed quasi-analytical approach made possible to  develop a theory of the FMR excitation of finite dot arrays, taking into account the edge effects. It was also shown that, using the developed theory, it is possible to calculate the absorption spectra of finite dot  arrays using  much less computation effort than required for the direct micromagnetic simulations of a similar system. Moreover, the quasi-analytical representation of the partial magnetic susceptibility tensors provides a way to understand how each spin wave mode of the array interacts with the applied electromagnetic driving field.

  We have also developed a computer program implementing our method~\cite{bib:Lisenkov:program} and illustrated the application of the developed  theory on the examples that fall into two categories: edge effects and domain-wall effects. In particular, it has been demonstrated that the FMR absorption spectrum calculated using our quasi-analytical theory for a  finite array of magnetic nanodots having a shape of right triangle agrees very well with the results obtained by direct micromagnetic simulations.
  An example of an array with a complex elementary cell containing two different magnetic dots was also considered. It was demonstrated that such arrays, while having reciprocal spectrum of bulk spin wave modes, may have a non-reciprocal spectrum of the edge spin wave modes.

  The spectra of spin wave excitations localized at  the domain walls in arrays of magnetic nanodots were calculated for two cases: a domain wall between the two ferromagnetic static magnetization states and a domain wall between the two chessboard antiferromagnetic static magnetization states. In the first case, it was demonstrated that the modes localized on the domain-wall have a very high degree of non-reciprocity. The propagation losses for these modes appear to be significantly different for the opposite propagation directions, which can be used for the development of nano-sized microwave isolators and circulators. For the case of an array existing in a chessboard antiferromagnetic static magnetization state with domain walls (which is a natural state for a demagnetized array) it has been shown  that the spin wave modes localized on the domain walls may produce a significant contribution to the FMR absorption spectra. In contrast to the spin wave modes localized at the array's edges, that have been considered in previous examples, the contribution to the FMR response by the modes localized on the domain wall in the chessboard antiferromagnetic ground state depends on the size of the domains, providing a tool to estimate the size of the domains, and, therefore, the ``quality'' of the static magnetization  state of the dot array.

\begin{acknowledgments}
This work was supported by the the Center for NanoFerroic Devices (CNFD) and the Nanoelectronics Research Initiative (NRI), by the DARPA MTO/MESO grant “Coherent Information Transduction between Photons,Magnons, and Electric Charge Carriers”, by the U.S. Army Contract from TARDEC, RDECOM, and by the Grant No. 1305574 from the National Science Foundation of the USA.
I.L. and S.N. acknowledge Russian Scientific Foundation, Grant \#14-19-00760 for financial support. I.L. thanks Mr. Steven Louis for the proofreading of the manuscript.
\end{acknowledgments}

\appendix
\section{An extension of the theory beyond the macrospin approximation}
\label{app:ext}
In the theory devolved above we used  the same demagnetization tensor $\N(\vec{r})$ for static~\eqref{eq:form:sum_static} and dynamic~\eqref{eq:form:sum_dynamic} equations. This approximation is correct, when the dots are assumed to be magnetized uniformly among their volume (``macrospin approximation'') and the crystalline anisotropy is uniaxial. Here we show, that the developed theory can be extended to  cover more general situations.

We demonstrate below that the theory can be substantially generalized if we use different static and dynamic demagnetization tensors in~\eqref{eq:form:sum_static} and \eqref{eq:form:omega_general}:
\begin{equation}
    B_i\vec{\mu}_i = \Bext_i - \mu_0 M_s\sum_j\op{N}^\text{st}_{ij}\cdot\vec{\mu}_j,\label{eq:app:sum_static}
    \tag{\ref{eq:form:sum_static}$'$}
\end{equation}
and
\begin{equation}
\op\Omega_{ij} = \gamma B_i\delta_{ij}\op I + \gamma \mu_0 M_s \op N^\text{dyn}_{ij},
\tag{\ref{eq:form:omega_general}$'$}\label{eq:app:omega_general}
\end{equation}
where  $\op{N}^\text{st}_{ij}$ and  $\op N^\text{dyn}_{ij}$ are the static and dynamic demagnetization tensors found in each particular case. In all the equations of the above developed theoretical formalism the static and dynamic demagnetization enter independently, so it is easy to trace all the changes caused by using different demagnetization tensors for static and dynamic magnetization.
\subsection{High-order crystalline anisotropy}
A high-order crystalline anisotropy is usually rather weak, compared to the the uniaxial anisotropy of the magnetic material and the shape anisotropy  of the magnetic sample. However, in some particular cases the high-order anisotropy can be qualitative important.  For example, this happens when the shape and crystalline anisotropies are not present ( e.g. when magnetic elements are spherical and made of an isotropic material~\cite{bib:Gurevich:1996,bib:Kaczer:1974}) or when the shape and crystalline anisotropies cancel each other.

As an example of the high-order anisotropy here we consider the first-order cubic anisotropy. Let us assume that the energy of the cubic anisotropy is $K_{c1}$ and the crystalline axes are: $\vec{e}_1$, $\vec{e}_{2}$ and $\vec{e}_{3}$. The  energy added by the first-order cubic anisotropy to the energy of a single magnetic dot can be written~\cite{bib:Stancil:2009} as (here and below we drop the dot's index):
\begin{equation}
    W_c = \dfrac{K_{c1}}{M_s^4}\sum_{\alpha\beta\gamma\delta}T_{\alpha\beta\gamma\delta} M_\alpha M_\beta M_\gamma M_\delta,
\end{equation}
where
\begin{multline}
    T_{\alpha\beta\gamma\delta} = \vec{e}_1\otimes\vec{e}_1\otimes\vec{e}_2\otimes\vec{e}_2+\\
\vec{e}_1\otimes\vec{e}_1\otimes\vec{e}_3\otimes\vec{e}_3+
\vec{e}_2\otimes\vec{e}_2\otimes\vec{e}_3\otimes\vec{e}_3.
\end{multline}
Thus, the effective field caused by the anisotropy can be found as:
\begin{equation}
    \vec{B}_{c1} = -\dfrac{\partial W_{c1}}{\partial \vec{M}} = -
    \dfrac{4K_{c1}}{M_s^4}\sum_{\alpha\beta\gamma}T^\text{sym}_{\alpha\beta\gamma\delta}
 M_\alpha M_\beta M_\gamma,
    \label{eq:app:Bc1}
\end{equation}
where
\begin{equation}
    T^\text{sym}_{\alpha\beta\gamma\delta} = \dfrac{1}{24}\left(T_{\alpha\beta\gamma\delta} +  T_{\beta\alpha\gamma\delta} + T_{\beta\gamma\alpha\delta} + \dots\right)
\end{equation}
To linearize~\eqref{eq:app:Bc1} we use expansion~\eqref{eq:form:M_expansion}. In contrast with the uniaxial case, the static and dynamic magnetic fields are different. The static field can be found as:
\begin{equation}
    \vec{B}_{c1}^\text{st} = -\dfrac{4K_{c1}}{M_s}
\sum_{\alpha\beta\gamma}T^\text{sym}_{\alpha\beta\gamma\delta} \mu_\alpha \mu_\beta \mu_\gamma,
\end{equation}
while the dynamic field is found as:
\begin{equation}
    \vec{B}_{c1}^\text{dyn} = -\dfrac{4K_{c1}}{M_s} \op{N}_{c1}(\vec{\mu})\cdot\vec{m},
\end{equation}
with
\begin{multline}
    \op{N}_{c1} = \sum_{\alpha\beta\gamma}T^\text{sym}_{\alpha\beta\gamma\delta} (\mu_\alpha \mu_\beta + \mu_\alpha\mu_\gamma + \mu_\beta\mu_\gamma) =\\
    3 \sum_{\alpha\beta}T^\text{sym}_{\alpha\beta\gamma\delta} \mu_\alpha \mu_\beta.
\end{multline}
Thus, the correction to the demagnetization tensor caused by the cubic anisotropy can be written as:
\begin{gather}
    \op{N}^\text{st}_{ij} = \op{N}_{ij} + \delta_{ij} \dfrac{4K_{c1}}{3\muz M_s^2}\op{N}_{c1}(\vec{\mu}),\\
    \op{N}^\text{dyn}_{ij} = \op{N}_{ij} + \delta_{ij} \dfrac{4K_{c1}}{\muz M_s^2}\op{N}_{c1}(\vec{\mu}).
\end{gather}
Note, that the effective demagnetization tensor now explicitly depends on the magnetic ground state of a dot. The dynamical equations do not get any additional complexity, however the static problem~\eqref{eq:app:sum_static} becomes considerably more complicated. Although, in a particular case, when the static component is aligned along one of the axes of the cubic anisotropy $\vec{e}_3=\vec{\mu}$ the correction to the demagnetization tensor has a simple form~\cite{bib:Stancil:2009}:
\begin{equation}
    \op{N}_{c1} = (\vec{e}_1\otimes\vec{e}_1 + \vec{e}_2\otimes\vec{e}_2)/2.
\end{equation}

\subsection{Dots with non-uniform dynamic magnetization }
For the sake of simplicity we derived the above presented theory implying that the magnetic dots have a uniform magnetization profile. Nevertheless, the theory can be extended for the cases, when the spatial profile of the spin wave mode is not uniform. The straightforward extension is possible when: (i)~the ground state of static magnetization of the dot is uniform; (ii)~the interactions between the dots do not alter the ground state of a single dot, e.g. when the inter-dot interaction is weaker than the exchange and dipolar self-interactions in a single dot; and (iii)~the eigenmodes of a dot are substantially separated in frequency. Under these assumptions, our approach remains sufficiently simple to allow the analytical analysis, and can describe dynamics of realistic experimental systems~\cite{bib:Kruglyak:2010,bib:Tacchi:2010,bib:Tacchi:2011}.

Let us assume that we know from either a micromagnetic simulation~\cite{bib:Vansteenkiste:2014, bib:Buijnsters:2014, bib:Grimsditch:2004} or from an analytical solution~\cite{bib:Jorzick:2002,bib:Grimsditch:2004, bib:Zivieri:2006} the linear magnetization dynamics of a single isolated magnetic dot in a form:
\begin{equation}
    \vec{M}(t, \vec{r})/\Ms = \vec{\mu} + \sum_{\lambda=1}^\mathfrak{L}  \vec{m}_\lambda(\vec{r})e^{-i\omega_\lambda t} + \text{c.c.},
    \label{eq:app:modes}
\end{equation}
where $\vec{\mu}$ is the static magnetization component, $\vec{m}_\nu(\vec{r})$ is the vector mode profile of the $\lambda$-th spin wave mode and $\mathfrak{L}$ is the number of these spin wave modes. For simplicity we assume that the modes do not interact. It is convenient to switch to a temporary coordinate system (denoted by the symbol $'$), in which $\vec\mu=\vec{z}'$, and write the distribution of the dynamic part of the magnetization in this system as~\cite{bib:Kalinikos:1986}:
\begin{equation}
    c_\lambda\vec{m}'_\lambda(\vec{r}) = \op{A}'_\lambda(\vec{r})\cdot\begin{pmatrix}1\\ i \\0\end{pmatrix},
\end{equation}
where $\op{A}'_\lambda(\vec{r})$ is a real dimensionless matrix-density:
\begin{equation}
    \op{A}'_\lambda=c_\lambda\begin{pmatrix}
                  \Real(m_x(\vec{r})& -\Imag(m_x(\vec{r})) & 0\\
                  \Imag(m_y(\vec{r})& \Real(m_y(\vec{r})) & 0\\
                  0&0& 0
            \end{pmatrix}.
\end{equation}
The normalization coefficient $c_\lambda$ is found from the condition:
\begin{equation}
    \dfrac{1}{V}\int_V \op{A}_\lambda(\vec{r})\cdot\op{J}\cdot\op{A}_\lambda(\vec{r}) d^3\vec{r} = \op{J},
\end{equation}
where $V$ is the volume of the dot and the operator $\op J$ is defined in the same way as in~\eqref{eq:form:main_dynamic}.

Following the same procedure as in [\onlinecite{bib:Beleggia:2003}], we obtain the Fourier representation of the effective demagnetization tensor:
\begin{equation}
    \op{N}_{\lambda, \vec{K}} = \dfrac{1}{V K^2}\op{A}_{\lambda, \vec{K}}\cdot (\vec{K}\otimes\vec{K})\cdot\op{A}^*_{\lambda, \vec{K}},
\end{equation}
where $\vec{K}$ in the three-dimensional reciprocal vector and $\op{A}_{\lambda,\vec{K}}$ is:
\begin{equation}
    \op{A}_{\lambda, \vec{K}} = \int_V \op{A}_\lambda(\vec{r}) e^{-i\vec{K}\cdot\vec{r}}d^3\vec{r}.
\end{equation}

After that, the effective dynamic demagnetization tensor $\op{N}^\text{dyn}_{\lambda, ij}$ and its in-plane Fourier image $\op{N}^\text{dyn}_{\lambda, \vec{k}}$ can be found straightforwardly~\cite{bib:Verba:2012}:
\begin{equation}
    \op{N}^\text{dyn}_{\lambda, \vec{k}} = \dfrac{1}{2\pi} \int \op{N}_{\lambda, \vec{k} +\vec{z}\kappa_z} d\kappa_z.
\end{equation}

Now one substitutes this tensor into~\eqref{eq:app:omega_general} and finds eigen-frequencies $\omega_\nu$ and spin-wave mode profiles $\vec{m}_{\nu, n}$ of the collective spin-wave excitations in an array of interacting nanodots using~\eqref{eq:mult:edge_dynamic}. After that, the distribution of magnetization within a single dot for the $\nu$-th \emph{array} mode can be found for each dot in the array:
\begin{equation}
    \vec{m}_{\nu,\lambda, n}(\vec{r}) = \op{A}_\lambda(\vec{r})\cdot\vec{m}_{\nu, n}.
\end{equation}

A case of multiple \emph{interacting} modes within a dot can be considered introducing a new ``layer'' of multi-vectors, representing amplitudes of different modes, similarly to the approach used for a complex unit cell in this paper. The operators $\op{J}$ will, however, have to be renormalized, as the different modes may carry a different magnetic moment.

\section{Multi-vector algebra}
\label{app:mva}
``Multi-vectors'' are introduced as the first rank objects in a multi-vector space. Instead of scalars for ``ordinary'' Euclidean vectors, each multi-vector ($\mvec{a}$) is an ordered collection of three-dimensional vectors $\vec{a}_i$. A multi-vector of the size $P$ contains $P$ three-dimensional vectors:
\begin{equation}
    \mvec{a} =
    \begin{pmatrix}\vec{a}_{1}\\\vec{a}_{2}\\\vdots\\\vec{a}_P\end{pmatrix}.\label{eq:app:multi-vector}
\end{equation}
Below we define the algebra in the multi-vector space. First of all we define a bilinear operation (scalar product) of two multi-vectors of the size $P$:
\begin{equation}
    \mvec{a} \cdot \mvec{b} =
    \begin{pmatrix}\vec{a}_{1}\\\vec{a}_{2}\\\vdots\\\vec{a}_P\end{pmatrix}\cdot
    \begin{pmatrix}\vec{b}_{1}\\\vec{b}_{2}\\\vdots\\\vec{b}_P\end{pmatrix}=
    \sum_{i=1}^P\vec{a}_{i}\cdot\vec{b}_{j}.
    \label{eq:app:dot}
\end{equation}
The second rank objects in the multi-vector space are the block-matrices $P$ by $P$, each element of which is a 3 by 3 matrix of scalars. A product of a multi-vector and a block-matrix is defined as follows:
\begin{multline}
\mop{M}\cdot\mvec{a} =
\begin{pmatrix}
    \op{M}_{11} & \cdots & \op{M}_{1P} \\
    \vdots & \ddots & \vdots\\
    \op{M}_{P1} & \cdots & \op{M}_{PP}
\end{pmatrix}\cdot
\begin{pmatrix}\vec{a}_{1}\\\vdots\\\vec{a}_P\end{pmatrix}=\\
\begin{pmatrix}\sum_{i=1}^P\op{M}_{1i}\cdot\vec{a}_i\\\vdots\\
               \sum_{i=1}^P\op{M}_{Pi}\cdot\vec{a}_i\end{pmatrix} = \mvec{b},
\label{eq:app:matvec}
\end{multline}
and returns another multi-vector. Although it is never used in the paper, one can also define
scaling a multi-vector as a multiplication of a multi-vector by a scalar and other higher rank operations.

The multi-vectors are convent for analytical and numerical calculations. In computer algebra systems the operations~\eqref{eq:app:dot} and~\eqref{eq:app:matvec} can be defined as custom user operations and one can analyze the object in the multi-vector space analytically.

For numerical simulations one can formally use multi-vectors as ordinary vectors of scalars of the size $3P$:
\begin{equation}
    \mvec{a}=
    \begin{pmatrix}a^x_1\\a^y_1\\a^z_1\\\vdots\\a^x_P\\a^y_P\\a^z_P\end{pmatrix}\cdot
\end{equation}
One can easily check, that the operations~\eqref{eq:app:dot} and~\eqref{eq:app:matvec} remain valid in this case.

It is important to stress, that our notation for ``multi-vectors'' is not related to the ``Multivector Calculus'' described in~\cite{bib:Hestenes:1968}.

\bibliography{em_cc}
\end{document}